\documentclass[12pt]{iopart}
\usepackage{graphicx}
\usepackage{iopams}

\begin{document}

\title{Recipes for spin-based quantum computing}

\author{Veronica Cerletti, W. A. Coish, Oliver Gywat and Daniel Loss}

\address{Department of Physics and Astronomy, University of Basel,
  Klingelbergstrasse 82, 4056 Basel, Switzerland}

\begin{abstract}
Technological growth in the electronics industry has historically been measured
by the number of transistors that can be crammed onto a single microchip.
Unfortunately, all good things must come to an end; spectacular growth in the
number of transistors on a chip requires spectacular reduction of the transistor
size.  For electrons in semiconductors, the laws of quantum mechanics take over
at the nanometre scale, and the conventional wisdom for progress (transistor
cramming) must be abandoned.  This realization has stimulated extensive research
on ways to exploit the \emph{spin} (in addition to the orbital) degree of
freedom of the electron, giving birth to the field of \emph{spintronics}.
Perhaps the most ambitious goal of spintronics is to realize complete control
over the quantum mechanical nature of the relevant spins.  This prospect has
motivated a race to design and build a spintronic device capable of complete
control over its quantum mechanical state, and ultimately, performing
computations: a quantum computer.

In this tutorial we summarize past and very recent developments which point the
way to spin-based quantum computing in the solid-state.  After introducing a set
of basic requirements for any quantum computer proposal, we offer a brief
summary of some of the many theoretical proposals for solid-state quantum
computers.  We then focus on the Loss-DiVincenzo proposal
for quantum computing with the spins of electrons confined to quantum dots.
There are many obstacles to building such a quantum device.  We address these,
and survey recent theoretical, and then experimental progress in the field.  To
conclude the tutorial, we list some as-yet unrealized experiments,
which would be crucial for the development of a quantum-dot quantum computer. 
\end{abstract}


\section{Introduction}

The fields of semiconductor physics and electronics have been successfully
combined for many years. The invention of the transistor meant a revolution for
electronics and has led to significant development of semiconductor physics and
its industry. More recently, the use of the spin degree of freedom of electrons,
as well as the charge, has attracted great
interest~\cite{awschalom:2002a,dassarma:2004a}. In addition to applications for
spin electronics (spintronics) in conventional devices, for instance based on
the giant magneto-resistance effect~\cite{baibich:1998a} and spin-polarized
field-effect transistors~\cite{datta:1990a}, there are applications that exploit
the quantum coherence of the spin. This was encouraged by ground-breaking
experiments that showed coherent spin transport
over long distances in semiconductors and long electron-spin dephasing times, on
the order of 100 nanoseconds~\cite{kikkawa:1998a,kikkawa:1998b}. In addition,
spin-polarized carrier injection from magnetic to non-magnetic semiconductors
has been demonstrated~\cite{fiederling:1999a,ohno:1999a}. Since the electron
spin is a two-level system, it is a natural candidate for the realization of a
quantum bit (qubit)~\cite{loss:1998a}. A qubit is the basic unit of information
in quantum computation, a discipline which attempts to radically improve the
performance of computers by exploiting the quantum properties of the system used
as hardware. The confinement of electrons in semiconductor structures like
quantum dots allows for better control and isolation of the electron spin from
its environment. Control and isolation are important issues to consider for the design
of a quantum computer.
 

The field of quantum computing was born in the 80's, motivated by the
miniaturization of electronic devices. Moore's-law predictions on the
exponential growth of the transistor density in microchips raises the question
of the possible future direction of the electronics industry.  In
particular, since small ($\sim$nm) systems are governed by the laws of quantum
mechanics, nanoscale hardware components must show quantum behaviour. Since
computers are built-up from these electronic components, this leads to the idea
of quantum computers. A different approach brought Richard Feynman to the
concept of quantum computing~\cite{feynman:1985a}. The simulation of the
dynamics of quantum systems on conventional computers is a hard task, meaning
that the computational resources needed to simulate a quantum system increase
exponentially with its size; the states of a quantum system are represented as
elements in a Hilbert space, and therefore, the dimension of the space needed to
describe the state of $n$ qubits is $2^n$. Thus, the simulation of $n$ qubits
requires an exponential number ($2^n$) of classical bits. Feynman's idea was
that this problem could be solved by simulating the object of study
with a system of the same nature; in this case it implies the use of a quantum
device to simulate a quantum system.


The efficient solution of problems that were previously considered intractable
has aroused some interest.  Researchers have begun thinking about how to exploit
the quantum properties of a system to perform calculations.  Following early
work by David Deutsch on the power of universal quantum computing
\cite{deutsch:1985a}, one of the first practical quantum algorithms was
presented by David Deutsch and Richard Jozsa in 1992~\cite{deutsch:1992a}. The
problem it solves is very simple (it determines
whether a function is constant or balanced), but it showed for the first time an
advantage in using quantum mechanics for computing. In 1994 quantum computation
captured world-wide attention, as Peter Shor presented his quantum algorithm for
the prime factorization of integers~\cite{shor:1994a}. This had a significant
impact due to the striking advantage of this algorithm with respect to its
classical counterparts. The time required to factor a number $N$ on a classical
computer with currently-known algorithms grows exponentially with the number of
digits $\log N$ ($\propto e^{(\log N)^\alpha}$), while in the case of Shor's
algorithm, the growth is bounded by a polynomial ($\propto (\log N)^\alpha$) \cite{shor:1997a}. In addition to
this fundamental breakthrough in computational complexity, Shor's algorithm also has potential practical
relevance; the difficulty of the factorization problem is the key to the
security of cryptographic codes. These codes include the RSA (Rivest, Shamir,
and Adelman) encryption scheme, widely used in the Internet, at banks and in
secret services. Nevertheless, there is still no formal proof that Shor's
algorithm outperforms any potential classical algorithm. This is different from
the case of another quantum algorithm created by Lov Grover in
1997~\cite{grover:1997a}, which shows a definite improvement over the classical
case, although the speed-up is less impressive than for Shor's
algorithm. Grover's algorithm is designed to perform a search in an unsorted
database. The time required to find one desired element out of $N$ is
proportional to $\sqrt{N}$, while in the classical case it is proportional to
$N$.  At the same time that the first quantum algorithms were proposed, the
first quantum error correcting codes were developed. The possibility to
implement error correcting codes encouraged researchers to work on physical
implementations of quantum computers, since these codes relax the demands on
control over noise and undesired interactions of the computer with the surrounding
environment.  Since the
appearance of the first quantum algorithms, quantum computation has undergone a
rapid development and growth, both from the theoretical and applied points of
view. Many setups have been proposed for the hardware of a quantum
computer~\cite{loss:1998a,gershenfeld:1997a,kane:1998a,cirac:1995a,shnirman:1997a},
arising from different fields of research including cold trapped atoms, nuclear
magnetic resonance, Josephson junctions, and electrons in quantum dots, just to
mention a few. The sort of physical attributes exploited in each case for the
representation, storage and manipulation of information varies over a wide
range.

Formally, a quantum computation is performed through a set of transformations,
called {\em gates}~\cite{preskill}. A gate applies a unitary transformation $U$
to a set of qubits in a quantum state $|\Psi\rangle$. At the end of the
calculation, a measurement is performed on the qubits (which are in the state
$|\Psi'\rangle=U\,|\Psi\rangle$). There are many ways to choose sets of {\em
universal} quantum gates. These are sets of gates from which any computation can
be constructed, or at least approximated as precisely as desired. Such a set
allows one to perform any arbitrary calculation without inventing a new gate
each time. The implementation of a set of universal gates is therefore of
crucial importance. It can be shown that it is possible to construct such a set
with gates that act only on one or two qubits at a time~\cite{barenco:1995a}.

The successful implementation of a quantum computer demands that some basic
requirements be fulfilled. These are known as the DiVincenzo
criteria~\cite{divincenzo:2000a} and can be summarized in the following:

\begin{enumerate}
\item{{\em Information storage\---the qubit:} We need to find some {\em quantum}
property of a {\em scalable} physical system in which to encode our bit of
information, that lives long enough to enable us to perform computations.}
\item{{\em Initial state preparation:} It should be possible to set the state of
the qubits to 0 before each new computation.}
\item{{\em Isolation:} The quantum nature of the qubits should be tenable; this
will require enough isolation of the qubit from the environment to reduce the
effects of decoherence.}
\item{{\em Gate implementation:} We need to be able to manipulate the states of
individual qubits with reasonable precision, as well as to induce interactions
between them in a controlled way, so that the implementation of gates is
possible. Also, the gate operation time $\tau_s$ has to be much shorter than the
decoherence time $T_2$, so that $\tau_s/T_2\ll r$, where $r$ is the maximum tolerable
error rate for quantum error correction schemes to be effective.}
\item{{\em Readout:} It must be possible to measure the final state of our
qubits once the computation is finished, to obtain the output of the
computation.}
\end{enumerate}

To construct quantum computers of practical use, we emphasize that the {\em
scalability} of the device should not be overlooked. This means it should be
possible to enlarge the device to contain many qubits, while still adhering to
all requirements described above. It should be mentioned here that this
represents a challenging issue in most of the physical setups proposed so
far. In this respect, very promising schemes for quantum computation are the
proposals based on solid-state
qubits~\cite{loss:1998a,kane:1998a,shnirman:1997a,averin:1998a,privman:1998a,imamoglu:1999a,meier:2003a,levy:2001a,ladd:2002a,friesen:2003a,stoneham:2003a},
which could take advantage of existing technology. In the following, we will
concentrate on proposals based on solid-state qubits, describing them in more
detail and summarizing recent achievements in the field.





\section{Proposals for quantum computing}

Before even the most rudimentary quantum circuits can be built, the elementary
registers (qubits) and quantum gates must be designed.  If any proposed design
is to be considered for experiment, it should first be subjected to a battery of 
theoretical tests to ensure its feasibility in real physical situations.  The 
five DiVincenzo criteria \cite{divincenzo:2000a} (introduced in section 1)
provide a simple checklist for the basic requirements of any physically 
realizable quantum computer.   Demonstrating strong adherence to these criteria 
is a daunting task, which requires a broad understanding of material 
properties, physical phenomenology and the quantum mechanical time evolution of 
these systems.  To make matters worse, a quantum computer, by \emph{necessity}, must 
remain in a phase-coherent state far from thermodynamic equilibrium under conditions of
strong time-dependent inter-qubit interactions (required for gating 
operations).  These conditions are beyond the reach of much
of the theoretical physicist's toolbox and therefore make the
development of new proposals both a challenging and exciting endeavour.

The first proposals for quantum computing made use of cavity quantum
electrodynamics (QED) \cite{domokos:1995a}, trapped ions \cite{cirac:1995a}, and
nuclear magnetic resonance (NMR) \cite{gershenfeld:1997a}.  All of these
proposals benefit from potentially long decoherence times, relative to their
respective gating times (however, see section \ref{sec:hybrid} below for a
discussion of the relative decoherence times in trapped ion systems that have
been realized in experiment).  In all three cases, this
is due to a very weak coupling of the qubits to their environment.  The long
decoherence times for these proposals and existing experimental expertise led to
quick success in achieving experimental realizations.  A conditional phase gate
was demonstrated early-on in cavity-QED systems \cite{turchette:1995a}. The
two-qubit controlled-{\sc not} gate, which, along with single-qubit rotations
allows for universal quantum computation \cite{barenco:1995a} has been realized
in single-ion \cite{monroe:1995a} and two-ion \cite{schmidt-kaler:2003a}
versions.  The most remarkable realization of the power of quantum computing to
date is the implementation of Shor's algorithm \cite{shor:1994a} to factor the
number 15 in a liquid-state NMR quantum computer \cite{vandersypen:2001a}.  In
spite of their great successes, the proposals based on cavity-QED, trapped ions
and NMR may not satisfy the first DiVincenzo criterion.  Specifically, these
proposals may not meet the requirement that the quantum computer can be
scaled-up to contain a large number of qubits \cite{loss:1998a}.  The
requirement for scalability motivated the Loss-DiVincenzo proposal
\cite{loss:1998a} for a solid-state quantum computer based on electron spin
qubits. This proposal was quickly followed by a series of proposals for
alternate solid-state realizations \cite{shnirman:1997a, averin:1998a,
privman:1998a, kane:1998a, imamoglu:1999a, meier:2003a, levy:2001a, ladd:2002a,
friesen:2003a, stoneham:2003a} and realizations for trapped atoms in optical
lattices that may also be scalable \cite{brennen:1999a, duan:2003a}.  In the
following sections we give a non-exhaustive survey of some of these proposals.
The goal of this survey is to demonstrate how the various requirements for
quantum computing have been met through example.

\subsection{\label{sec:Loss-DiVincenzo}Quantum dot quantum computing}
\begin{figure}
\begin{center}
\scalebox{0.2}{\includegraphics{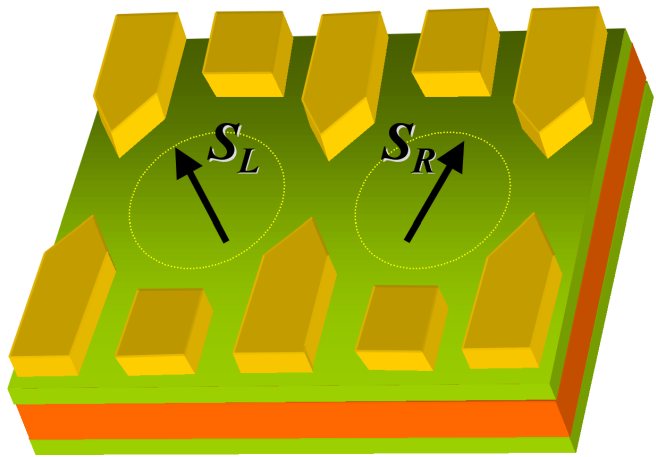}}
\end{center}
\caption{\label{fig:twoqubitexchange} Two neighbouring electron spins confined
  to quantum dots, as in the Loss-DiVincenzo proposal.  The lateral confinement
  is controlled by top gates.  A time-dependent Heisenberg exchange 
  coupling $J(t)$ can be pulsed high by pushing the electron spins closer, 
  generating an appreciable overlap between the neighbouring orbital wave
  functions.} 
\end{figure}

\begin{figure}
\begin{center}
\scalebox{0.5}{\includegraphics{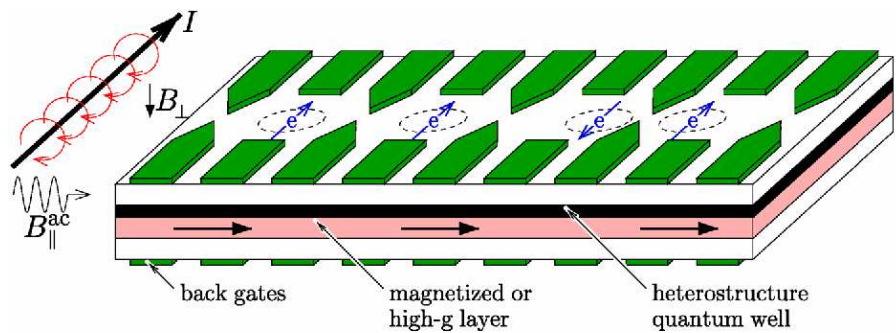}}
\end{center}
\caption{\label{fig:qdarray} An array of exchange-coupled quantum dots.  Top
  gates provide lateral confinement and allow pulsing of the exchange
  interaction for two-qubit operations (in this image the two dots on the left
  are decoupled, whereas the two dots on the right are coupled). Back gates
  could pull electrons down into a region of higher $g$-factor to allow
  single-qubit operations in conjunction with applied constant ($B_\perp$) and
  rf ($B^\mathrm{ac}_\parallel$) magnetic fields.}
\end{figure}

The qubits of the Loss-DiVincenzo quantum computer are formed from the two spin
states ($\left|\uparrow\right>,\left|\downarrow\right>$) of a confined electron.
The considerations discussed in this proposal are generally applicable to
electrons confined to any structure, such as atoms, molecules, etc., although
the original proposal focuses on electrons localized in quantum dots.  These
dots are typically generated from a two-dimensional electron gas (2DEG), in
which the electrons are strongly confined in the vertical direction.  Lateral
confinement is provided by electrostatic top gates, which push the electrons
into small localized regions of the 2DEG (see figures \ref{fig:twoqubitexchange}
and \ref{fig:qdarray}). Alternative quantum-dot structures are discussed in
section 4.  Initialization of the quantum computer can be achieved
by allowing all spins to reach their thermodynamic ground state at low
temperature $T$ in an applied magnetic field $B$ (i.e., virtually all spins will
be aligned if the condition $\left|g\mu_\mathrm{B}B\right| \gg k_\mathrm{B}T$ is
satisfied, with $g$-factor $g$, Bohr magneton $\mu_\mathrm{B}$, and Boltzmann's
constant $k_\mathrm{B}$).  Several alternative initialization schemes have been
investigated (see sections \ref{sub:Spin-Initialization-and} and 
\ref{sec:optical initialization}). Single-qubit
operations can be performed, in principle, by changing the local effective
Zeeman interaction at each dot individually.  To do this may require large
magnetic field gradients \cite{wu:2004a}, $g$-factor engineering
\cite{medeiros-ribeiro:2003a}, magnetic layers (see figure \ref{fig:qdarray}), the inclusion of
nearby ferromagnetic dots \cite{loss:1998a}, polarized nuclear
spins, or optical schemes (see section \ref{sub:Single-Qubit-Rotations}).
In the Loss-DiVincenzo proposal, two-qubit operations are performed by pulsing the electrostatic
barrier between neighbouring spins.  When the barrier is high, the spins are
decoupled.  When the inter-dot barrier is pulsed low, an appreciable overlap develops
between the two electron wave functions, resulting in a non-zero Heisenberg
exchange coupling $J$.  The Hamiltonian describing this time-dependent process
is given by 
\begin{equation} 
\label{eqnExchangeHamiltonian}
H(t) = J(t)\mathbf{S}_L\cdot\mathbf{S}_R.
\end{equation} 
This Hamiltonian induces a unitary evolution given by the
operator $U=\mathcal{T}\exp\left\{-i\int H(t) dt/\hbar \right\}$, where
$\mathcal{T}$ is the time-ordering operator.  If the exchange is pulsed on for a
time $\tau_\mathrm{s}$ such that $\int J(t) dt/\hbar=J_0\tau_s/\hbar=\pi$, the states of
the two spins, with associated operators $\mathbf{S}_L$ and $\mathbf{S}_R$, as
shown in figure \ref{fig:twoqubitexchange}, will be exchanged.  This is the {\sc
swap} operation.  Pulsing the exchange for the shorter time $\tau_\mathrm{s}/2$ generates
the ``square-root of {\sc swap}'' operation, which can be used in conjunction
with single-qubit operations to generate the controlled-{\sc not} (quantum {\sc
xor}) gate \cite{loss:1998a}.  In addition to the time scale $\tau_\mathrm{s}$, which
gives the time to perform a two-qubit operation, there is a time scale
associated with the rise/fall-time of the exchange $J(t)$.  This is the switching time
$\tau_\mathrm{sw}$.  When the relevant two-spin Hamiltonian takes the form of an ideal
(isotropic) exchange, as given in (\ref{eqnExchangeHamiltonian}), the total spin
is conserved while switching.  However, to avoid leakage to higher
\emph{orbital} states during gate operation,
the exchange coupling must be switched adiabatically.  More precisely,
$\tau_\mathrm{sw} \gg 1/\omega_0\approx10^{-12}\,\mathrm{s}$, where
$\hbar\omega_0\approx1$\,meV is the energy gap to the next orbital state
\cite{loss:1998a, schliemann:2001a, schliemann:2002b, requist:2004a}.  We stress
that this time scale is valid only for the ideal case of a purely isotropic exchange
interaction.  When the exchange interaction is anisotropic, different
spin states may mix and the relevant time
scale for adiabatic switching may be significantly longer.   
For scalability, and application of quantum error correction
procedures in \emph{any} quantum computing proposal, it is important to turn off
inter-qubit interactions in the idle state.  In the Loss-DiVincenzo proposal,
this is achieved with exponential accuracy since the overlap of neighbouring
electron wave functions is exponentially suppressed with increasing separation. A
detailed investigation of decoherence during gating due to a bosonic environment
was performed in the original work of Loss and DiVincenzo.  Since then, there
have been many studies of leakage and decoherence in the context of the
quantum-dot quantum computing proposal.  We discuss some of these studies in
section \ref{sec:Obstacles}, after reviewing alternative solid-state proposals
for quantum computing.

\subsection{Superconducting qubits}
Among the first proposals for solid-state quantum computing were qubits based on
superconducting Josephson junctions \cite{mooij:1999a, orlando:1999a,
  averin:1998a, shnirman:1997a, makhlin:1999a}.  These
proposals were quick to take advantage of the macroscopic quantum coherence
afforded in such structures, and a large and well-developed literature on their
non-equilibrium dynamics \cite{schoen:1990a}.  The development of new designs
for superconducting qubits has become an industry unto itself.  There are, for
example, designs that exploit the $d$-wave pairing symmetry of cuprate
high-temperature superconductors \cite{ioffe:1999a,zagoskin:2002a} and Andreev bound
states \cite{chtchelkatchev:2003a}.  The observation of coherent oscillations in
superconducting qubits \cite{vion:2002a,yang:2002a} was a watershed for the field
of solid-state quantum information, demonstrating conclusively that quantum
coherence could be generated and sustained for many precession periods
($\sim10^4$ in the experiment by Vion \emph{et al.} \cite{vion:2002a}). More recent achievements of
the superconducting proposals include the demonstration of a controlled-{\sc not}
gate \cite{yamamoto:2003a} and the controlled coupling
of a superconducting qubit to a single microwave photon mode
\cite{wallraff:2004a}.  In spite of these successes, the reduced visibility of
coherent oscillations and the particular sources and nature of decoherence for
these devices remains the subject of investigations \cite{burkard:2004a,
tian:2002a, meier:2004b}. Extensive reviews of Josephson-junction qubits 
can be found in references \cite{makhlin:2001a,burkard:2004b}.

\subsection{Quantum computing and the quantum Hall effect}
\begin{figure}
\scalebox{0.7}{\includegraphics{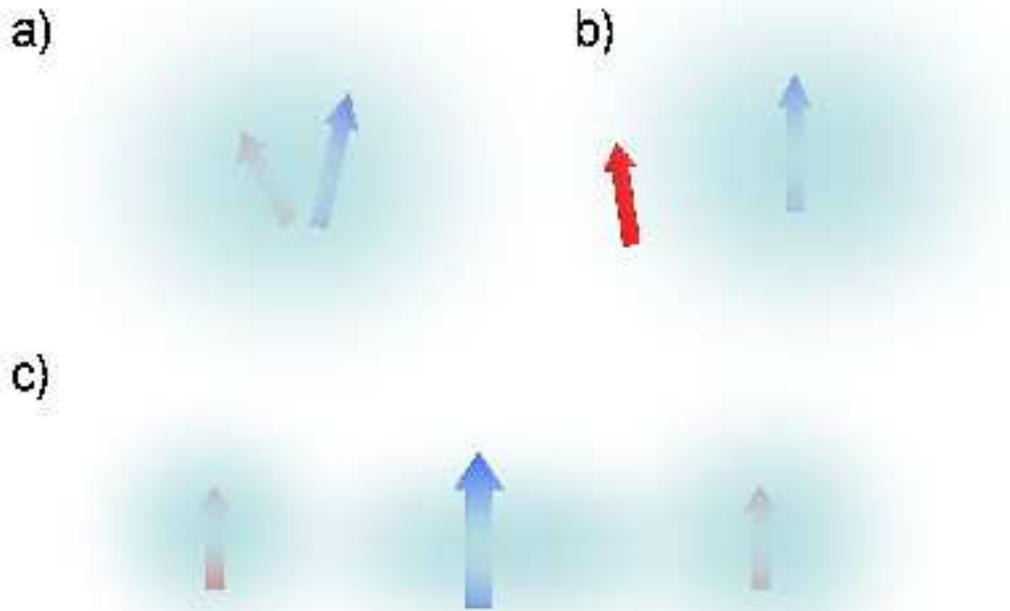}}
\caption{\label{fig:hyperfine}Schematic diagram illustrating the Fermi contact hyperfine
  interaction.  Electron spins are represented by longer arrows and nuclear
  spins are represented by shorter arrows.  The
  electron cloud is indicated with shading.  (a) The
  direct exchange interaction is proportional to the electron density at the position of
  the nucleus. The interaction is strong when the electron is close.  (b) The 
  interaction is weaker when the nuclear spin is far from the centre of the
  electron wave function. (c) When two nuclear spins couple to the same
  (delocalized) electron, an effective exchange interaction between nuclear
  spins is generated. }
\end{figure}

Based on observed long lifetimes for nuclear spin states, Privman {\it et
  al.} \cite{privman:1998a} have proposed a quantum computer composed of nuclear
spins embedded in a two-dimensional electron gas (2DEG) in the quantum-Hall
regime.  The qubits of their proposal are encoded in the states of nuclear
spins, which must be sufficiently separated to avoid dipolar coupling, but close
enough ($\sim 10\,\mathrm{nm}$) to allow significant interaction via the
electron gas.  Initialization of the qubits is achieved by placing
spin-polarized conducting strips with a current of electrons above the nuclear
spin qubits.  The contact hyperfine interaction between electron and nuclear
spins causes a polarization transfer from the electrons in the strips to the
nuclear spins, preferentially orienting the nuclear spins along the electron
spin polarization direction. Readout is performed in a complementary manner,
with a transfer of polarization from the nuclear spins to electrons in the
conducting strips.  Single-qubit operations are performed via
standard NMR pulses, which would require strong magnetic field gradients or many
different nuclear spin species to bring single specific nuclear spins into
resonance, while leaving the other qubits unchanged.  A pairwise interaction
between the nuclear spin qubits is necessary for the implementation of
two-qubit gates.  This interaction is generated by a superexchange, mediated by electrons in the
quantum Hall fluid that surrounds the nuclear spins (see
figure \ref{fig:hyperfine} (c)).  The electron gas that
couples the nuclear spins should be in the quantum Hall regime to avoid Friedel
oscillations in the electron density, and hence, a rapidly-varying RKKY
exchange \cite{vagner:1995a, mozyrsky:2001b}.  To perform computations, it is necessary
to switch the interaction on and off.  In the original work of Privman {\it et al.} it
was not clear how best to pulse the inter-qubit interaction \cite{privman:1998a}.
Topics such as switching error (leakage to states outside of the qubit basis)
and perhaps the most important of all, decoherence, are not addressed in the
original work of Privman {\it et al.}  However, subsequent studies of the decoherence of nuclear
spins in the integer quantum Hall regime have led to the prediction that the
decoherence time for these qubits could be as long as $T_2 \simeq
10^{-1}$\,s \cite{mozyrsky:2001b, mozyrsky:2001a}.

\subsection{Shallow-donor quantum computing}    
Following the proposals of Loss-DiVincenzo and Privman {\it et al.}, 
Kane \cite{kane:1998a} has introduced a proposal that
takes advantage of the long lifetimes of nuclear spins (as in the proposal of
Privman \emph{et al.}) and electrically-controlled gating of two-qubit interactions (as in
the Loss-DiVincenzo proposal).  This proposal uses the nuclear spins of $^{31}\mathrm{P}$ donor
impurities in silicon as its qubits.  Each donor impurity is associated with a
weakly-bound electron in an $s$-type orbital state.  One- and two-qubit
operations are performed with electrostatic ``$A$-gates" and ``$J$-gates",
respectively. These gates take their names from the conventional symbols for the
contact hyperfine ($A$) and spin exchange ($J$) coupling constants.  The
$A$-gates adjust the position of the electron cloud relative to the donor
nucleus (see figures. \ref{fig:hyperfine} (a), (b)).  In this way, the magnitude
of the contact hyperfine interaction is varied, bringing the nuclear spins in
and out of resonance with a uniform applied magnetic field.  Two-qubit
operations are performed via an electron-mediated superexchange between
neighbouring nuclear-spin qubits (figure \ref{fig:hyperfine} (c)), as in the
proposal of Privman {\it et al.} The $J$-gates adjust the overlap of electron
clouds on neighbouring impurities, thus controlling the strength of the
superexchange. Readout of the qubits is performed by transferring nuclear spin
information back to the electron spins and observing the resulting orbital
electron wave function via standard capacitive techniques. The original work of
Kane includes a discussion of decoherence due to a fluctuating gate
voltage. This work does not, however, discuss the influence of the nuclear
dipole-dipole interaction \cite{desousa:2003c}, problems associated with a
violently position-dependent exchange interaction \cite{koiller:2002a}, or
decoherence mechanisms that could affect the electron spin during gate operation
or measurement. These mechanisms include spin-orbit coupling and the contact
hyperfine interaction with surrounding nuclear spins.

\subsection{Spin-cluster qubits}
With the exception of proposals such as the ``exchange-only'' scheme
\cite{divincenzo:2000b}, nearly all quantum computing architectures require
single-qubit operations.  Addressing single spin-qubits with magnetic resonance
pulses usually requires magnetic field gradients or $g$-factor engineering to
bring the spins into resonance individually.  To implement two-qubit gates, the
spin qubits must typically be separated by very small distances (on the order of
the electron wave function: $\simeq 50$ nm in quantum dots, $\simeq 10$ nm in
the proposal of Privman {\it et al.}, and $\simeq 5$ nm for an electron bound to
a phosphorus donor in silicon).  This requirement leads to
extremely large magnetic field or $g$-factor gradients, which may not be
practical in a typical laboratory setting.  To resolve this issue, Meier {\it et
al.}  \cite{meier:2003a} have proposed a scheme for quantum computing based on
antiferromagnetic spin clusters, rather than single spins.  In this proposal the
quantum computer consists of many spin clusters.  Each cluster contains an odd
number of antiferromagnetically exchange-coupled spins.  The two basis states of
the qubit are encoded in the ground-state doublet formed by two total-$S_z$
eigenstates for one cluster.  Since its basis corresponds to two total
spin-$S_z$ eigenstates with an associated magnetic moment, the qubit can be
manipulated with a magnetic field to perform single-qubit operations in the same
way as for a single spin-1/2.  Furthermore, the qubit basis is protected from
higher-lying states by a gap of order $\Delta\sim J\pi^2/2n_c$ for a cluster
containing $n_c$ spins with exchange coupling $J$ \cite{meier:2003a}.  To
perform two-qubit operations, separate clusters are coupled at their ends by a
tunable exchange.  Initialization of the qubits is achieved by cooling the
system to its ground state in a strong magnetic field, as in the Loss-DiVincenzo
proposal.  Since the two orthogonal states of the ground-state doublet resemble
classical N{\'e}el ordering with the magnetization alternating
$\uparrow\downarrow\uparrow\ldots$, or $\downarrow\uparrow\downarrow\ldots$,
readout can be performed, in principle, with a local magnetization measurement.
Decoherence due to magnetic field fluctuations has been considered in this work.
There is no increase in the decoherence rate (over the single-spin rate) for a
magnetic field that fluctuates uniformly over the cluster, although there is a
linear increase with cluster size for local magnetic fields that fluctuate
independently.

\subsection{\label{sec:molmagqc}Quantum computing with molecular magnets}

Recently there has been significant interest in using molecular magnets for
quantum computing applications. These systems exhibit a number of interesting
quantum-mechanical features that can be probed in experiment, including quantum
tunnelling \cite{friedman:1996a}, interference effects \cite{wernsdorfer:1999a},
and the coherent superposition of high-spin states \cite{debarco:2004a} (for a
review, see reference \cite{leuenberger:2003c}).
Additionally, molecular magnets can be well-understood in terms of relatively simple spin
Hamiltonians, which means that high-resolution spin resonance spectroscopy
\cite{hill:2003a} or specific heat measurements \cite{troiani:2004a} can be used
to extract the relevant coupling constants empirically.  

Leuenberger and Loss \cite{leuenberger:2001a} have introduced a proposal to
perform Grover's algorithm in ensembles of large-spin molecular magnets. Since
this proposal relies only on a well-defined multilevel quantum system with
non-equidistant level spacing, the same procedure can be applied to nuclear
spins in GaAs in the presence of the nuclear quadrupole interaction
\cite{leuenberger:2002a, leuenberger:2003b} or to multilevel Josephson junction
devices, where coherent oscillations have now been observed
\cite{claudon:2004a}. We note that Grover's algorithm has been implemented
experimentally using atomic Rydberg states \cite{ahn:2000a}.  While these
proposals and experiments are valuable for demonstrating the practical
implementation of quantum computation, they rely on single multilevel systems,
and are therefore not scalable. Finally, the very recent proposal of Troiani
{\it et al.} \cite{troiani:2004a,troiani:2004b} suggests using the molecular magnet
$\mathrm{Cr}_7\mathrm{Ni}$ as a real-world implementation of the spin-cluster
quantum computing scheme discussed in the previous section.

\subsection{\label{sec:silicon}Silicon valley}
For quantum computing and spintronics applications, silicon has advantages over
other semiconductors.  First, silicon has long been a staple for the electronics
industry. Second, the spin-orbit interaction in silicon is weak (evidence of this is provided by the
small difference in effective electron-spin $g$-factor from the free value).  Third,
natural silicon contains only $4.7\%$ nuclear-spin-carrying isotopes, which
significantly reduces the effects of the contact hyperfine interaction relative
to materials such as (Ga/In)As.  Silicon quantum dots are, however, not as
advanced as the alternatives made from III-V semiconductors, and silicon is an
indirect gap semiconductor (in contrast to the direct gap material GaAs), which
limits its use in optical applications. Nevertheless, silicon's prevalence in industry means
that purification and fabrication techniques are usually more well-established
than for other semiconductors.

Levy \cite{levy:2001a} has suggested specializing the Loss-DiVincenzo proposal
to Ge/Si quantum dots.  Instead of using top-gates to confine electron spins laterally, these
dots would be defined by patterning a ferroelectric material (which has a finite
electric dipole moment) on the surface of a 2DEG.  In this proposal, two-qubit
gating operations would be performed by applying optical excitation to the
ferroelectric, which changes the local electric field that defines neighbouring
quantum dots.  This change in the local electrostatic potential generates a pulsed
exchange interaction between neighbouring electron spins.  The electrical
pulsing, which defines the rise-time (switching time) $\tau_\mathrm{sw}$ for the
exchange coupling occurs at terahertz frequencies
($\tau_\mathrm{sw}\approx10^{-12}\,\mathrm{s}$).  This short time scale will
likely violate the adiabaticity criterion discussed in section
\ref{sec:Loss-DiVincenzo}.  To satisfy the adiabaticity criterion, Levy suggests
using a third dot to mediate a superexchange between qubit dots, as in
\cite{recher:2000b}.  

Ladd {\it et al.} \cite{ladd:2002a,abe:2003a} have proposed an all-silicon
quantum computer, where the qubits are generated from $^{29}\mathrm{Si}$ nuclear
spins embedded in a $^{28}\mathrm{Si}$ matrix.  In a sufficiently large magnetic
field gradient, provided by a strong Dy ferromagnet, single-qubit operations
could be performed with NMR pulses and two-qubit operations could be performed
by pulsing the dipole-dipole interaction between neighbouring nuclear spins
(which would be suppressed in the idle state with an appropriate sequence of NMR
averaging pulses).  Readout in this proposal would be provided by magnetic
resonance force microscopy (MRFM) \cite{rugar:1992a,berman:2000a}, where the
nuclear spin state couples to vibrational modes of a cantilever or thin silicon
bridge.  Recent success in the detection of the \emph{existence} of a single
electron spin with MRFM is very promising, although it has not yet been shown
experimentally that a single-spin quantum \emph{state} can be measured using this
technique \cite{rugar:2004a} (see also Section \ref{sec:spindetection} below for a
description of MRFM detection).

The recent proposal of Friesen {\it et al.} \cite{friesen:2003a} uses electron
spins confined to silicon quantum dots.  This proposal is based on the
Loss-DiVincenzo quantum dot quantum computer, specialized to a silicon
environment.  Friesen {\it et al.}  have developed a strategy for
initialization and readout via spin-charge conversion, which has been modelled in
detail in reference \cite{friesen:2004a}.  Two-qubit operations are performed,
as in the original Loss-DiVincenzo proposal, by pulsing a direct exchange
between neighbouring electrons using electrostatic gates to increase or decrease
the overlap between neighbouring electron wave functions.  Friesen {\it et al.}
have performed a detailed calculation of exchange vs. gate voltages to find the
correct operating regime for their proposed quantum computer.  In addition, they
consider decoherence due to fluctuations in gate voltage, but do not address
other channels of decoherence.

Perhaps one of the most challenging quantum computing proposals comes from
Stoneham {\it et al.}  \cite{stoneham:2003a,stoneham:2003b,rodriquez:2004a}. The
qubits of their proposal consist of electron spins bound to deep-donor
impurities in silicon.  Between each pair of qubits, there is a control atom. By
optically exciting an electron from the highest valence state of the control
atom to a molecular state formed between the deep donors, a superexchange is
generated between neighbouring qubits, which can be turned off again by
stimulated de-excitation. The qubits in this proposal are addressed individually
by using ``site selectivity'' (every qubit has a unique environment, therefore a
unique energy-level structure).  Since the energies involved in the gating
process are large, Stoneham {\it et al.} suggest that this proposal could
potentially operate at room temperature.

\subsection{\label{sec:hybrid}Hybrid proposals}

In an attempt to extract the best from both worlds, there have been proposals
for hybrid quantum computers.  These proposals aim to couple ideas from 
proven approaches to quantum computing (cavity QED, trapped ions and trapped
atoms) with the benefits offered by solid-state implementations.  

Imamo\=glu \emph{et al.} \cite{imamoglu:1999a} have suggested a scheme that
combines cavity-QED and spin-based quantum dot quantum computing.  The qubits of
this proposal are encoded in the spin states of quantum dots, as in the
Loss-DiVincenzo proposal.  The quantum dots are contained within a semiconductor
microcavity, with well-defined optical modes.  Single-qubit operations are
performed by addressing individual dots with optical fibres and coupling the
spin-up and spin-down states via a Raman process, induced by laser excitation.
To perform two-qubit operations, distant electron spins are coupled via a
delocalized cavity mode.  This induces an XY-like interaction between electron
spins.  In the original work of Imamo\=glu \emph{et al.}, it was shown that an
XY-interaction and single-qubit rotations are sufficient to perform a two-qubit 
{\sc CNOT}-gate.  Single-spin readout could be performed in this proposal by
exciting a spin-selective transition in which a photon is emitted (or not
emitted) depending on the electron spin state.  In this way, the state of the
single electron spin is determined by the presence or absence of a single photon.

Quantum optical proposals and implementations often use the hyperfine (spin) and
vibrational states of trapped ions and atoms as their qubits.  The coupling
strengths for these states are typically very small relative to their
solid-state counterparts. This means that decoherence times ($T_2$) for these
implementations are relatively long (for example, $T_2\simeq170\,\mu$s in
reference \cite{demarco:2002a}).  For the same reason, however, the relevant
gating times ($\tau_s$) are also relatively long ($\tau_s\simeq10\,\mu$s for a
{\sc CNOT} gate in reference \cite{demarco:2002a}). The ratio of gating to decoherence time that
has been observed $r=\tau_s/T_2\approx1/17$ greatly exceeds current
estimates for the error threshold allowable for effective quantum error 
correction.  To remedy this potential difficulty, a very recent proposal by Tian
{\it et al.} \cite{tian:2004a} suggests a combined quantum optical and
solid-state device.  In this proposal the states of trapped atoms or
ions would be used as a long-lived quantum information storage device during the
idle state. When fast one- or two-qubit operations are to be performed,
information is transferred to some solid-state device (electron spins in quantum
dots or superconducting qubits) then returned again to the storage device when the
operation is complete.
\section{\label{sec:Obstacles}Obstacles to quantum dot quantum computing}
Several major obstacles to quantum dot quantum computation were identified and
addressed in the original work of Loss and DiVincenzo \cite{loss:1998a}, and
later elaborated upon \cite{divincenzo:1999a, burkard:1999a, awschalom:2002a}.  These obstacles
include entanglement (the creation and transport of a coherent superposition of
states), gating error (leakage to higher states outside of the qubit basis
during gate operation), and perhaps most importantly, coherence (the
preservation of any given superposition in the presence of a coupling to the
environment).  In the rest of this section we review work that has been done
to understand and possibly surmount these three obstacles in the context of the
Loss-DiVincenzo proposal.

\subsection{Flying qubits and entanglement generation}

In addition to the five DiVincenzo criteria for quantum \emph{computation} introduced
in section 1, there are two ``desiderata", which are important for
performing quantum \emph{communication} tasks.  These desiderata, which were
addressed in \cite{divincenzo:1999a}, are summarized in the following
statements \cite{divincenzo:2000a}:

\begin{enumerate}
\addtocounter{enumi}{5}
\item{The ability to inter-convert stationary and flying qubits.}

\item{The ability to faithfully transmit flying qubits between distant locations.}
\end{enumerate}

The whimsical term ``flying qubits" refers to qubits that can be conveniently 
moved from place to place.  The most obvious (and common) choice for a flying
qubit is provided by the polarization states of
photons \cite{turchette:1995a}. In the context of quantum-dot quantum computing,
this has led to a number of proposals for the conversion of quantum information \cite{imamoglu:1999a, gywat:2002a,
leuenberger:2003a, leuenberger:2004a, troiani:2003a} or entanglement \cite{cerletti:2004a}
from spin to light, and \emph{vice versa}.  More recent work has
suggested that ``free electron quantum computation" may be possible in
principle \cite{beenakker:2004a, stace:2004a}, in which mobile electrons (in some material)
travelling between dots could replace photons as the flying qubit medium of choice.

Deeply connected to the implementation of flying qubits is the creation of
nonlocal entanglement.  The race to create and measure
\cite{divincenzo:1999a,burkard:2000a,loss:2000a,egues:2002a,burkard:2003a,samuelsson:2004b} entangled particle
pairs has led to a virtual industry of so-called ``entangler" proposals for the
spin \cite{divincenzo:1999a,burkard:2000a,choi:2000a,recher:2001a, lesovik:2001a, melin:2001a, costa:2001a, oliver:2002a,
bose:2002a, recher:2002a, gywat:2002a, bena:2002a, saraga:2003a, bouchiat:2003a,
recher:2003a, saraga:2004a} and orbital
\cite{samuelsson:2003a,beenakker:2003a,samuelsson:2004a} degrees of freedom.
These proposals have the very ambitious goal of generating and spatially
separating a many-particle quantum superposition that can not be factorized into
single-particle states. The canonical example of such a state for the spin
degree of freedom is the singlet formed from two spin-1/2 particles:
$(\left|\uparrow\downarrow\right>-\left|\downarrow\uparrow\right>)/\sqrt{2}$.
The various efforts related to spin entanglement include proposals to extract
and separate spin-singlet pairs from a superconductor through two quantum dots
\cite{recher:2001a} or Luttinger-liquid leads \cite{recher:2002a,bena:2002a} and
proposals that generate entanglement near a magnetic impurity
\cite{costa:2001a}, through a single dot \cite{oliver:2002a}, from biexcitons in
double quantum dots \cite{gywat:2002a}, through a triple dot
\cite{saraga:2003a}, and from Coulomb scattering in a two-dimensional electron
gas \cite{saraga:2004a}.  Entanglement generation and measurement remains a
lofty goal for those working on solid-state quantum computing, theorists and
experimentalists alike. Recent experiments \cite{zumbuhl:2004a} that have
measured the concurrence (an entanglement measure) for electrons in the ground
state of a two-electron quantum dot point to a promising future for
entanglement-related phenomena in the solid state (see also section
\ref{sub:Two-Qubit-Gates}).  For recent reviews on entanglement generation and
measurement, see references \cite{egues:2003a, recher:2004a}.

\subsection{Gating error}
Hu and Das Sarma have evaluated the probability for double-occupancy of one of
the dots in the Loss-DiVincenzo proposal using Hartree-Fock and molecular
orbital techniques \cite{hu:2000a}.  They suggest that it may be difficult to
achieve both a significant exchange coupling and low double-occupancy
probability.  Schliemann {\it et al.} \cite{schliemann:2001a, schliemann:2002b}
and more recently Requist {\it et al.} \cite{requist:2004a} have
investigated the probability for double-occupancy gating errors in a pair of
coupled quantum dots during {\sc swap} gate operation.  Through numerical and
analytical study they have found that the Loss-DiVincenzo proposal is very
robust against double-occupancy errors when operated in the adiabatic regime (defined in section
\ref{sec:Loss-DiVincenzo}).  Barrett and Barnes \cite{barrett:2002a} have
subsequently shown that \emph{orbital} dephasing can result in a significant
error rate ($10^{-2}$--$10^{-3}$ errors per gate operation).  This is comparable to current estimates for the maximum error rate allowable for
quantum error correction to be effective \cite{steane:2003a}, but further studies on the nature of
the spin-orbit interaction have suggested that the spin-orbit coupling can be
minimized with careful pulsing of the exchange during gate operations (see
section \ref{subsub:Spin-orbit}).  When the potential barrier between quantum
dots is pulsed low, the overlap between nearest-neighbour dots is appreciable,
while that between next-nearest and next-next-nearest neighbours is
exponentially suppressed with distance.  In spite of the smallness of these
interactions, Mizel and Lidar \cite{mizel:2004a} have recently suggested that
three- and four-spin interactions in a realistic quantum computing proposal may
lead to substantial gating errors.  These problems are, however, specific to a
particular architecture, and it is possible that they could be corrected or exploited
by adjusting the device design \cite{mizel:2004a}.

\subsection{Decoherence}

Every experimental apparatus shows some small fluctuations in electrostatic
voltage and applied magnetic field.  These fluctuations, acting on an electron
spin in a quantum dot, will inevitably induce decay of the spin directly through
the Zeeman interaction (in the case of a fluctuating magnetic field), or
indirectly through spin-orbit coupling (in the case of a fluctuating electric
field).  The effect of these fluctuations can be treated accurately (for a weak
coupling to the electron spin) by the phenomenological spin-boson model within a
Born-Markov approximation, as derived in reference \cite{loss:1998a}.  The coupling
of the electron spin to the bath cannot always be treated as weak, and effects
of the bath memory (non-Markovian evolution) may be important for achieving the
level of accuracy required to perform quantum error correction.  For these
reasons, the solution to this model has recently been extended to obtain
non-Markovian effects \cite{loss:2003a} and corrections beyond the Born
approximation \cite{divincenzo:2004a} in the case of ohmic dissipation in the bath.

\begin{table}
\begin{tabular} {|c|c|c|c|c|}\hline
& Symbol & Description & Estimate & reference\\
\hline\hline
1. & $\hbar\omega_0$ & Size-quantization energy & 1\,meV &  \cite{burkard:1999a}\\
\hline
2. & $J$ & Electron spin exchange coupling & $10^{-1}$\,meV &  \cite{burkard:1999a}\\
\hline
3. & $\hbar\max\{|\alpha|,|\beta|\}/l $ & Spin-orbit coupling strength & $10^{-2}$\,meV &  \cite{miller:2003a}\\
\hline
4. & $A$ & Hyperfine interaction (polarized nuclei) & $10^{-1}$\,meV &
 \cite{paget:1977a}\\
\hline
5. & $A/\sqrt{N}$ & Hyperfine interaction (unpolarized nuclei) &
$10^{-4}$\,meV & --\\
\hline
6. & $A/N$ & Knight shift dispersion &
$10^{-6}$\,meV & --\\
\hline
7. & $N\mu_\mathrm{B}\mu_N/l^3$ & Electron-nuclear dipolar coupling &
$10^{-7}$\,meV & --\\
\hline
8. & $\hbar/\tau_\mathrm{dd}$ & Nuclear-nuclear dipolar coupling & $10^{-8}$\,meV &  \cite{paget:1977a}\\
\hline 
9. & $\mu_\mathrm{B}^2/l^3$ & Electron-electron dipolar coupling & $10^{-9}$\,meV &
--\\
\hline 
\end{tabular}
\caption{\label{tab:energyscales}Relevant energy scales for the Loss-DiVincenzo quantum computing
  proposal.  The above estimates are based on a GaAs dot of lateral size 
  $l=30\,\mathrm{nm}$ containing $N=10^5$ nuclear spins. The typical
  size-quantization energy $\hbar\omega_0$ and exchange coupling $J$ for a dot
  of this size are taken from reference \cite{burkard:1999a}.  The Rashba ($\alpha$)
  and Dresselhaus ($\beta$) coefficients were extracted from experimental data in
  reference \cite{miller:2003a}.  The hyperfine interaction constant $A$ was
  estimated from a weighted average over the hyperfine coupling constants for
  the three nuclear spin species in GaAs in reference \cite{paget:1977a}.  The
  nuclear spin dipolar coupling is estimated from the linewidth of the NMR
  resonance in \cite{paget:1977a}, which gives a correlation time
  $\tau_\mathrm{dd}\approx10^{-4} s$.}  
\end{table}


Fluctuations in voltage and magnetic field are artifacts of a given experimental
apparatus.  In principle, these fluctuations can be reduced with improved
electronics, and can therefore be regarded as \emph{extrinsic} sources of decoherence.
In addition to these extrinsic sources, there are sources of decoherence that
are \emph{intrinsic} to the quantum dot qubit design.  These include the coupling of
the electronic spin to phonons in the surrounding lattice or other fluctuations
via the spin-orbit interaction \cite{divincenzo:1998a, burkard:1999a,
khaetskii:2000a, khaetskii:2001a, levitov:2003a, glavin:2003a,
cheng:2004a,golovach:2004a} and coupling of the electron spin to surrounding
nuclear spins via the contact hyperfine interaction \cite{burkard:1999a,
erlingsson:2001a, erlingsson:2002a, khaetskii:2002a, khaetskii:2003a,
schliemann:2002a, merkulov:2002a, desousa:2003a, desousa:2003b, marquardt:2004a,
erlingsson:2004a, coish:2004a, desousa:2004b, schliemann:2003a, vagner:2004a, 
yuzbashyan:2004a}.  A detailed understanding of the electron spin evolution
under the influence of these interactions is of fundamental interest and is necessary to implement reliable
quantum dot quantum computation.  The first step to understanding any
decoherence mechanism is to estimate its size.  In table \ref{tab:energyscales}
we give estimates for various energy scales related to decoherence and qubit
operation in the Loss-DiVincenzo proposal.

\subsubsection{\label{subsub:Spin-orbit}Spin-orbit coupling}
We would like to assess the spin-orbit coupling strength for typical quantum
dots.  Performing the standard non-relativistic expansion and reduction to a
two-component spinor for a Dirac electron to leading order in $1/m c^2$ leads to
the spin-orbit coupling term \cite{elliott:1954a} 
\begin{equation}
H_{\mathrm{so}} = \frac{\hbar}{2 m^2 c^2} \left(\nabla V(\mathbf{r})\times\mathbf{P}\right)\cdot\mathbf{S}. 
\end{equation}
In the above, $m$ is the electron mass, $c$ is the speed of light,
$V(\mathbf{r})$ is the potential experienced by the electron, $\mathbf{P}$ is
the momentum operator in three dimensions, and $\mathbf{S}$ is the electron spin-1/2 operator.
For a spherically symmetric parabolic confining potential,
$V(\mathbf{r})=m\omega_0^2r^2/2$, the spin-orbit coupling term is
$H_\mathrm{so}=(\omega_0^2/2 m c^2)\mathbf{L}\cdot\mathbf{S}$. Here, $\mathbf{L}=\mathbf{r}\times\mathbf{P}$ is the
orbital angular momentum operator, which can be substituted with $\hbar$ for
estimation purposes.  Comparing the strength of this coupling to the orbital
energy $\hbar\omega_0\approx 1\,\mathrm{meV}$ gives
$\left<H_{\mathrm{so}}\right>/\hbar\omega_0 \approx 10^{-7}$ \cite{divincenzo:1998a,
  burkard:1999a}.   This smallness of the spin-orbit coupling
compared to the orbital energy scale would suggest that the electron spin in quantum
dots is relatively free from external influences that couple to its charge.  In
realistic dots, however, the confining potential is neither smooth (it has a $1/r$
singularity at the centre of each lattice ion), nor spherically symmetric, and
the resulting spin-orbit interaction takes-on a more complicated form.  In a
crystalline solid, the spin-orbit interaction is the sum of structure inversion
asymmetry (Rashba) \cite{rashba:1960a} and bulk inversion asymmetry (Dresselhaus) 
\cite{dresselhaus:1955a} terms, which can be written for an electron confined 
to two dimensions as 
\begin{equation}
\label{eqn:HsoRashbaDresselhaus}
H_{\mathrm{so}} = \alpha(p_x \sigma_y - p_y \sigma_x) + \beta(-p_x \sigma_x +
p_y \sigma_y) + O\left(|\mathbf{p}|^3\right). 
\end{equation}
$\alpha\;(\beta)$ is the Rashba (Dresselhaus) coefficient,
$\mathbf{p}=(p_x,p_y)$ is the electron momentum operator in the $x$-$y$ plane,
and $\sigma_{x,y}$ are the usual Pauli matrices.  For a strongly two-dimensional
system, the cubic Dresselhaus term, of order $\sim |\mathbf{p}|^3$, can be
neglected relative to the Rashba and linear Dresselhaus terms, which have the
size $\sim p_{x,y}p_z^2$ \cite{dyakonov:1986a}.  In a two-dimensional quantum
dot, we replace $p_z\approx\hbar/d,p_{x,y}\approx\hbar/l$, where $d$ is the
2DEG thickness and $l$ is the lateral quantum dot size.  The cubic term is then
smaller than the linear Dresselhaus and Rashba terms by a
factor of order $\sim\left(d/l\right)^2$.  The Rashba and Dresselhaus
coefficients have been extracted from magnetoresistance data in a GaAs/AlGaAs
2DEG.  This gives the values $\hbar\beta = (4 \pm 1)\,\mathrm{meV}$-\AA\ and
$\hbar\alpha = (-5 \pm 1)\,\mathrm{meV}$-\AA\ \cite{miller:2003a}.  To estimate the
size of $H_{\mathrm{so}}$ given in (\ref{eqn:HsoRashbaDresselhaus}) for a
quantum dot containing a single electron, we replace the momenta by
$p_{x,y}\approx \hbar/l$, where $l = 10$--$100\,\mathrm{nm}$.  This gives 
the range $\left<H_{\mathrm{so}}\right> =
10^{-2}$--$10^{-1}\,\mathrm{meV}$. This estimate is significantly larger than
the value ($\simeq 10^{-7}$\,meV) for a simple parabolic confining potential.
All is not lost, however, since the spin-orbit coupling can only affect the spin
\emph{indirectly} through fluctuations in the orbital degree of freedom.  We can only
assess the real danger of this interaction through a correct microscopic
analysis of the spin-orbit Hamiltonian in the proper context.

The direct effects that a realistic spin-orbit interaction has on two-qubit
gating operations in a quantum computer have been explored by several authors.
Bonesteel {\it et al.} \cite{bonesteel:2001a} have shown that the effect of the
spin-orbit interaction on coupled quantum dot qubits can be minimized by using
time-symmetric qubit gating.  Subsequently, Burkard and Loss
\cite{burkard:2002a} have shown that the spin-orbit effect during gating can be
eliminated completely for appropriately chosen exchange pulse shapes (see also
reference \cite{stepanenko:2003a}). Additionally, there have been several
investigations into the possible spin-flip (relaxation) \cite{khaetskii:2000a,
khaetskii:2001a, levitov:2003a, glavin:2003a, cheng:2004a} and decoherence
\cite{semenov:2004a,golovach:2004a} mechanisms mediated by the spin-orbit
interaction and coupling to lattice phonons or other fluctuations.  In many
ways, an electron in the orbital ground state of a quantum dot is very similar
to an electron bound to a donor impurity site.  Since the spin relaxation and
decoherence times for electrons bound to shallow donors have been
investigated many years ago \cite{pines:1957a, klauder:1962a}, much of this work
has been used to accelerate progress for the analogous quantum dot structures.  

Khaetskii and Nazarov have calculated the rates for spin-flip transitions due to
the spin-orbit interaction both through direct relaxation from an excited
orbital state accompanied by a spin flip \cite{khaetskii:2000a}, and through a
virtual process between the two states of a Zeeman-split doublet within the same
orbital state \cite{khaetskii:2001a}.  The most effective spin-flip mechanism
for a transition between Zeeman-split states, which has a rate $1/T_1\propto
(g\mu_\mathrm{B}B)^5/(\hbar\omega_0)^4$, is significantly reduced for decreasing
magnetic field $B$ and increasing orbital energy $\hbar\omega_0$.

In the presence of spin-orbit coupling, a precessing spin induces an oscillating
electric field.  Levitov and Rashba \cite{levitov:2003a} have suggested that this
coupling may be a double-edged sword in view of applications to spintronics.  On
the positive side, the time-varying electric field might provide access to the
dynamics of a single isolated spin. The reverse mechanism, however, leads to
a further channel for spin relaxation from excitations in the dot leads.

There have been further studies of spin-lattice relaxation mechanisms that are
specialized to particular quantum dot architectures. Glavin and Kim \cite{glavin:2003a} have compared
results for Si quantum dots and donor impurities, and Cheng
{\it et al.} \cite{cheng:2004a} have performed a numerical exact-diagonalization study for GaAs quantum
dots, extending the validity of previous calculations to a more
realistic set of wave functions.  

The spin-flip (relaxation) time $T_1$ is important for applications of spintronics
involving classical information, encoded in the states $\left|\uparrow\right>$
and $\left|\downarrow\right>$.  However, for quantum computing tasks, the
relevant time scale is the spin decoherence time $T_2$, which is the
lifetime for a coherent superposition
$a\left|\uparrow\right>+b\left|\downarrow\right>$. Typically, the
decoherence time is much less than the relaxation time $(T_2 \ll T_1)$.  Golovach {\it et al.} have
shown that the fluctuations induced from spin-orbit coupling are purely
transverse to the direction of an applied magnetic field to leading order in the
coupling \cite{golovach:2004a}.  Because the fluctuations are purely transverse,
the corresponding $T_2$ time due to the combined spin-orbit and electron-phonon
interactions \emph{exceeds} the value of the longitudinal spin relaxation time,
giving $T_2=2T_1$.  Moreover, for phonons in three dimensions, the spectral
function is super-ohmic ($\sim\omega^3$) and thus the pure dephasing
contribution is absent, again ensuring that $T_2=2T_1$.  Provided other 
decoherence mechanisms can be arbitrarily suppressed, this result is very 
promising for applications of quantum dot quantum computing in view of recent
experiments that show exceptionally long $T_1$ times for single electron spins
confined to GaAs quantum dots (see section \ref{sub:Spin-Relaxation}).
      

\subsubsection{\label{sec:hyperfine}Spin-spin coupling} 
Unfortunately, the spin-orbit interaction is not the end of the decoherence
story.  The electron spin can also couple directly to other spins
embedded in the quantum computer device.  In a GaAs quantum dot, the electron
wave function contains approximately $N=10^5$ lattice nuclei, and 
every nucleus carries spin $I=3/2$.  The dominant
spin-spin coupling for this type of dot arises from the Fermi contact hyperfine interaction.
The Fermi contact hyperfine interaction for an electron with orbital envelope wave
function $\psi(\mathbf{r})$ and spin operator $\mathbf{S}$ interacting 
with surrounding nuclear spins $I_k$ is described by the spin Hamiltonian
\begin{equation}
H_\mathrm{hf} = \sum_k A_k\mathbf{S}\cdot\mathbf{I}_k;\;\;\;\;\;A_k=v_0 A|\psi(\mathbf{r}_k)|^2.
\end{equation}   
Here, $v_0$ is the volume of a crystal unit cell containing one nuclear spin.
Due to $H_\mathrm{hf}$, the electron spin will experience an effective magnetic
field (the \emph{Overhauser field}), which gives rise to an energy splitting on
the order of $p I A$, where $I$ is the total nuclear spin and $p$ is the nuclear
spin polarization.  For full polarization of the nuclear spin system, the
Overhauser field induces a splitting $\approx IA = 10^{-1}\,\mathrm{meV}$ in
GaAs.  In a typical unpolarized sample, we have $|p|\approx 1/\sqrt{N}$, which
gives a splitting $IA/\sqrt{N} \approx 10^{-4}\,\mathrm{meV}$ for a quantum
dot containing $N=10^5$ nuclear spins.  In addition, the nuclear spin at site
$k$ will experience an effective Zeeman splitting (\emph{Knight shift}) on the
order of $A_k$.  Since the coupling constants $A_k$ vary in space from $A_k
\approx A/N = 10^{-6}$\,meV near the dot centre to $A_k=0$ far from the dot,
nuclear spins at different sites will precess with different frequencies.  This
dispersion in the Knight shift will efficiently destroy collective states
generated in the nuclear spin system on a time scale $t\approx \hbar N/A \approx
1\,\mu\mathrm{s}$ \cite{deng:2004a}, and is therefore important for proposals
based on nuclear spin quantum computing.

In addition to the Fermi \emph{contact} hyperfine term, there is an \emph{anisotropic}
contribution to the hyperfine interaction.  For a widely separated electron and nucleus, the anisotropic hyperfine
interaction reduces to the interaction energy between point dipoles:
\begin{equation}
H_\mathrm{dd} = \sum_k\frac{(g\mu_\mathrm{B})(g_I\mu_N)}{r^3}\left\{
                \frac{3(\mathbf{I}_k\cdot\mathbf{r})(\mathbf{S}\cdot\mathbf{r})}{r^2}
                -\mathbf{I}_k\cdot\mathbf{S}\right\}.
\end{equation} 
For a microscopic derivation of $H_\mathrm{hf}$ and $H_\mathrm{dd}$, see
reference \cite{stoneham:1975a}.  If the electron spin is in a spherically
symmetric orbital $s$-state with the nuclear spin at its centre, the anisotropic
hyperfine interaction vanishes identically \cite{stoneham:1975a}.  The
contribution of this term from nuclear spins near the dot centre will therefore
be small, but for nuclear spins near the edge of the electron wave function,
which do not ``see'' a spherical electron spin distribution, it may become
appreciable.  Assuming approximately $N=10^5$ nuclear spins have a significant dipolar
coupling to the electron, we estimate the size of the electron-nuclear dipolar
interaction as $\left<H_\mathrm{dd}\right>\approx
N\mu_\mathrm{N}\mu_\mathrm{B}/l^3 \simeq 10^{-7}$\,meV, where $l = 30$\,nm
is the typical dot size.

The final spin-spin coupling directly associated with the electron is the
magnetic dipolar coupling of the electron to other electron spins in
neighbouring quantum dots.  This can be estimated as
$\mu_\mathrm{B}^2/l^3\approx 10^{-9}\,\mathrm{meV}$.  Although this coupling is
very weak for neighbouring single-electron quantum dots, it can become
significant at atomic length scales, and may be a significant source of
decoherence for other solid-state proposals \cite{meier:2003a}.  In addition to
direct electron spin coupling mechanisms, there are also significant mechanisms
that couple the environment to itself. For example, the nuclear
spins experience a mutual dipolar coupling.  This dipolar coupling causes the
nuclear environment to evolve dynamically, which can, in turn, affect the electron
through direct hyperfine coupling.  The nuclear spins evolve on a time scale
given by the dipolar correlation time $\tau_\mathrm{dd} = 10^{-4}$\,s.  The
time $\tau_\mathrm{dd}$ is determined from the linewidth of the NMR resonance (in bulk)
through $\hbar/\tau_\mathrm{dd} \simeq 10^{-8}$\,meV \cite{paget:1977a}.

There have been many studies of electron spin dynamics in the presence of the
strongest (Fermi contact hyperfine) spin-spin interaction.  Burkard {\it et al.}
\cite{burkard:1999a} showed that in the presence of the hyperfine interaction
with surrounding nuclear spins, the electron spin-flip transition probability
could be suppressed by applying a magnetic field $B$ or polarizing the nuclear
spin system (this probability is suppressed by the factor $1/p^2 N$ for $B=0$,
nuclear spin polarization $p$ and $N$ nuclear spins within the quantum dot).
Erlingsson {\it et al.} have investigated singlet-triplet transitions mediated
by the contact hyperfine interaction \cite{erlingsson:2001a} and transitions
between a Zeeman-split doublet \cite{erlingsson:2002a}.  In an investigation of
decoherence, Khaetskii {\it et al.} \cite{khaetskii:2002a,khaetskii:2003a} have
found an exact solution for the electron spin evolution under the action of
$H_\mathrm{hf}$ in the particular case of a fully-polarized nuclear spin system.
They found that only a small fraction ($\simeq 1/N$) of the electron spin
underwent decay and the resulting dynamics were described by a power-law or inverse
logarithmic decay at long times.  Schliemann {\it et al.}
\cite{schliemann:2002a} have performed exact diagonalizations on small nuclear
spin systems.  These exact diagonalization studies show that the hyperfine
interaction can be very efficient in causing decay of the electron spin in small
systems and that the dynamics of an ensemble are reproduced by the dynamics of a
randomly correlated initial nuclear spin state.  Yuzbashyan {\it et al.}
\cite{yuzbashyan:2004a} have recently found an exact closed-form solution 
for the classical (mean-field) analogue of this problem and highlighted its
connection to the dynamics of the BCS pairing model.

The gating operations performed on a quantum computer are performed on
\emph{single} isolated systems.  This raises the question of whether ensemble or
pure-state initial conditions should be used when calculating spin dynamics for
the purpose of quantum computing.  The free-induction decay of the
electron spin in the presence of an ensemble of nuclear spin configurations has
been investigated by Merkulov {\it et al.} \cite{merkulov:2002a}, who found a
rapid initial Gaussian decay of the electron spin with a time scale
$\tau\approx1\,\mathrm{ns}$ in GaAs.  Even for a \emph{single} quantum mechanical
initial state of the nuclear system, the electron-spin free-induction decay can
be severe. For a translationally-invariant direct-product nuclear
spin state with polarization $p$, and in the limit of a large number $N\gg1$ of
nuclear spins $I=1/2$, and large magnetic field $|g\mu_\mathrm{B}B| \gg A$, the
transverse electron spin $\left<S_+\right>_t=\left<S_x\right>_t+i\left<S_y\right>_t$
decays like a Gaussian \cite{coish:2004a} (up to a time-dependent phase factor):
\begin{equation} 
\label{eqn:gaussian-decay}
\left<S_+\right>_t \propto \left<S_+\right>_0e^{-\frac{t^2}{2\tau^2_c}};
\;\;\;\;\; \tau_c = \sqrt{\frac{N}{1-p^2}}\frac{2\hbar}{A}.  
\end{equation} 
In GaAs, and for polarization $p \approx 0$, we have $\tau_c \approx
5\,\mathrm{ns}$.  The time scale $\tau_c$ can be moderately extended by
polarizing the nuclear spin system.  However, even a polarization degree of
$99\%$ (the current record in a GaAs quantum dot is $60\%$ \cite{bracker:2004a},
and significant gate-controlled nuclear spin polarization has been seen in a
GaAs 2DEG in the quantum Hall regime \cite{smet:2002a}) would only 
extend the decay time by a factor of $10$.  If the state of the nuclear spins
could be prepared, e.g., via a measurement, in an eigenstate of the total
$z$-component of the nuclear Overhauser field, the decay in
(\ref{eqn:gaussian-decay}) would be removed.  Under these conditions, the electron
spin still undergoes a nontrivial non-Markovian (history dependent) dynamics on
a time scale given by the inverse Knight shift dispersion $\hbar
N/A\approx1\mu\mathrm{s}$.  This decay can be evaluated in the presence of a
sufficiently large magnetic field \cite{coish:2004a}.

An alternative way to remove the effects of the decay in (\ref{eqn:gaussian-decay})
is to perform a spin-echo sequence on the electron \cite{coish:2004a}.  The decay of the
Hahn spin-echo envelope due to spectral diffusion (which includes the effect
of the nuclear dipole-dipole interaction) has been investigated by de
Sousa {\it et al.} \cite{desousa:2003a,desousa:2003b} for a model with 
fluctuating classical nuclear spins $I=1/2$, that evolve in a Markovian
fashion.  This same model has recently been extended to larger nuclear spin
$I>1/2$ \cite{desousa:2004b}.

In addition to work on the time-dependent evolution of a localized electron
spin, there have been proposals for spintronic devices that use the contact
hyperfine interaction to their advantage.  These include a proposal for dynamic
polarization of nuclear spins via optical manipulation of localized electrons
\cite{imamoglu:2003a} and a proposal for a nuclear spin quantum memory
\cite{taylor:2003a, taylor:2003b, taylor:2004a} that takes advantage of
potentially long-lived nuclear states. The quantum memory proposal 
is limited by the Knight shift dispersion in quantum dots in the 
presence of an electron spin \cite{deng:2004a}. The electron must therefore be removed
from the dot after transferring quantum information to the nuclear spin.  In
this case, the nuclear spin state may live as long as the nuclear spin
dipole-dipole correlation time $\tau_\mathrm{dd}\approx10^{-4}\,\mathrm{s}$ (in GaAs) or possibly longer
if, for example, so-called WaHuHa NMR pulses are applied to suppress the dipole-dipole 
interaction \cite{taylor:2004a}.


\section{Experimental achievements}

In this section we present a selection of important experimental achievements
leading towards the implementation of quantum information processing
using electron spins in quantum dots.

\subsection{Single and coupled quantum dots\label{sub:Structures-of-Single}}
\begin{figure}[b]
\begin{center}
\scalebox{0.5}{\includegraphics{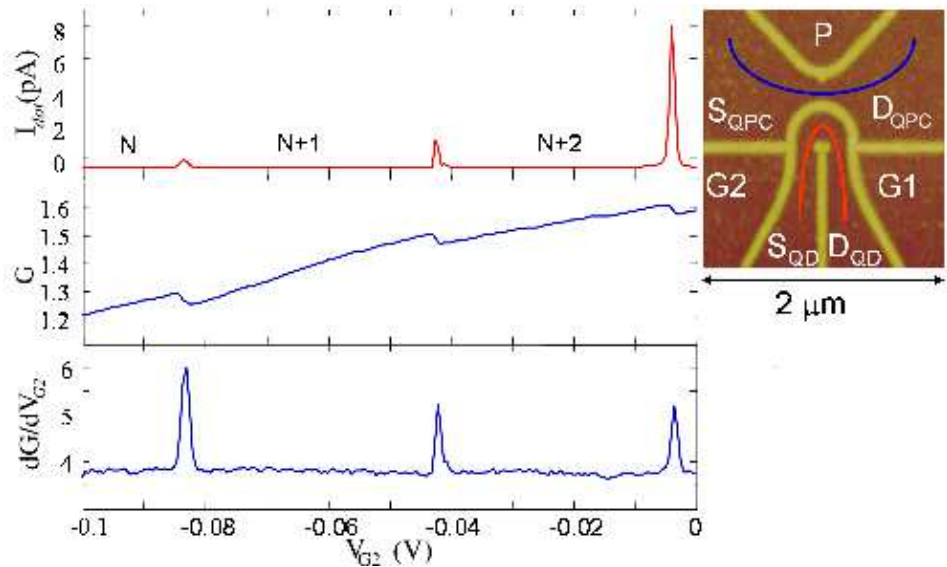}}
\end{center}
\caption{\label{fig:ensslinreadout}  Device (right) used to read-out the charge state of
a quantum dot with a quantum point contact (QPC) from reference \cite{schleser:2004a}.  The plunger gate ($P$)
controls the restriction of the QPC.  $S_\mathrm{QPC}$ and $D_\mathrm{QPC}$ are (respectively)
the source and drain of the QPC, and similarly, $S_\mathrm{QD}$ and $D_\mathrm{QD}$ are the
source and drain of the quantum dot.  The voltages associated with the gates
$G1$ and $G2$ can be varied to adjust the number $N$ of electrons on the dot
one-by-one. At left is a plot of the current through the dot $I_\mathrm{dot}$, and the d.c. and differential conductance through the QPC
(denoted by $G$ and $dG/dV_\mathrm{G2}$) as the gate voltage $V_\mathrm{G2}$ is
varied. (Figure courtesy of K Ensslin.) Reprinted with permission from Schleser R,
Ruh E, Ihn T, Ensslin K, Driscoll D C and Gossard A C 2004 {\it Appl. Phys.
Lett.} \textbf{85} 2005. \copyright 2004 American Institute of Physics.}
\end{figure}

We first discuss different experimental approaches to construct semiconductor
quantum dot structures that enable control over the spin degree of
freedom on the level of a single electron. The precise control of
the number of excess electrons in a quantum dot is a necessary prerequisite
to achieve control over the spin states of interest. The addition
of an electron from the surrounding material to a negatively charged
dot requires the charging energy $\delta\epsilon_{\mathrm{c}}$ to
overcome the electrostatic energy of other electrons in the dot. The
charging energy $\delta\epsilon_{\mathrm{c}}$ depends on the number
$N$ of charges confined in the dot. The regime (gate voltages) where the injection
of additional electrons into the dot is blocked due to 
$\delta\epsilon_{\mathrm{c}}$
is known as the Coulomb blockade regime (see figure 
\ref{fig:ensslinreadout}). In recent years,
a great deal of experimental effort has focused on the single-electron
regime ($N=1$) using different types of quantum dot structures. This
regime provides experimental access to a spin-$1/2$ in the dot. 
There are several possibilities to produce quantum dot structures
capable of confining single electrons.
The list of ingenious quantum dot production techniques has grown enormously
during the last years. Instead of presenting a complete list thereof,
we rather focus on a few techniques that have paved the way for the
first steps towards the implementation of quantum information processing
using spin states.

\begin{figure}[t]
\begin{center}
\includegraphics[%
      clip, 
      width=14cm, 
      keepaspectratio]{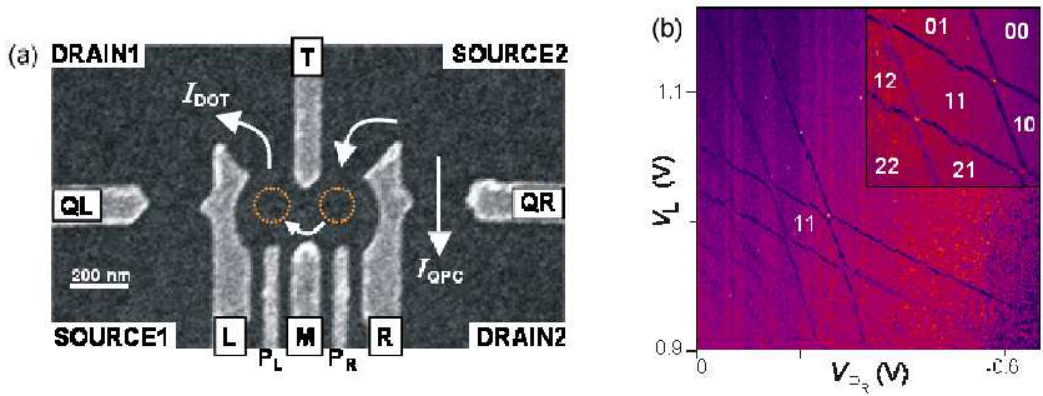}
\end{center}

\caption{(a) Scanning electron micrograph of a gated double dot structure 
with two adjacent quantum point contacts (QPCs). 
The circles indicate the dot positions. When a bias voltage is applied between 
source 2 and drain 1, a current $I_{\mathrm{DOT}}$ flows through the dots. 
Excess charges in the double dot modulate the current 
$I_{\mathrm{QPC}}$ through one of the QPCs \cite{field:1993a}. 
In this figure,  the current $I_{\mathrm{QPC}}$ flows from source 2 
to drain 2 and enables charge readout via the right QPC. 
(b) Charge stability (``honeycomb'') diagram  of the double 
quantum dot. 
The labels $nm$ indicate the regions (``Coulomb diamonds'') where 
$n$ ($m$) electrons are 
present in the left (right) dot. The colour scale indicates 
$dI_{\mathrm{QPC}}/dV_L$, measured as a function of the bias 
voltages $V_{\mathrm{L}}$ and $V_{\mathrm{\mathrm{P_R}}}$ applied to the gate
$\mathrm{L}$ and the 
right plunger gate $\mathrm{P_R}$, respectively. The inset shows a blow-up 
of the region around 11.  \label{Fig:doubledot}  
(Figure courtesy of L P Kouwenhoven.) Reprinted with permission from
Engel H-A, Kouwenhoven L P, Loss D and Marcus C M 2004 {\it Quantum Inf.
Process.} \textbf{3} 115. \copyright 2004 Springer-Verlag}
\end{figure}

\begin{figure}[t]
\vspace{2 cm}
\begin{center}
\includegraphics[width=8.6cm]{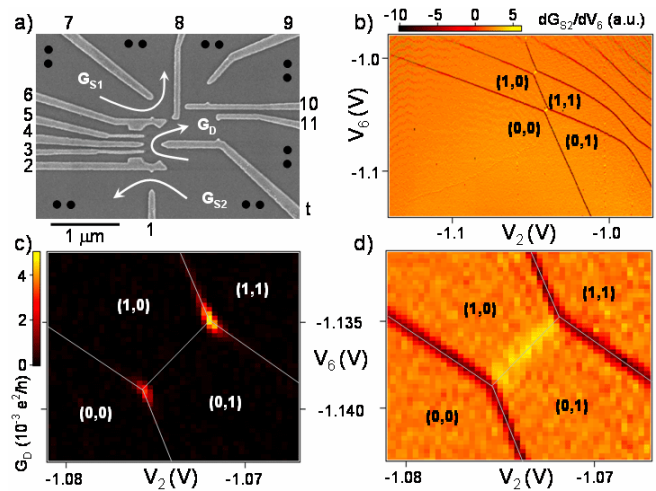}
\end{center}
\vspace{-1.5 cm} \caption{\label{fig: marcusdoubledot}(a) SEM micrograph of 
an electrically gated double quantum dot 
structure with neighbouring QPC charge detectors \cite{petta:2004a}. 
The symbols $\bullet$ denote ohmic contacts. (b) Large-scale
plot of the differential conductance $dG_{S2}$/$dV_{6}$ as a function 
of the voltages $V_{2}$ and $V_{6}$ applied to gates 2 and 6, 
respectively.
The number of electrons is indicated by $($$M$,$N$$)$, where $M$$($$N$$)$ is
the time-averaged number of electrons in the upper (lower) dot.
In (c) and (d), $G{_D}$  and $dG_{S2}$/$dV_{6}$  are shown, respectively, 
as a function of $V_{2}$ and $V_{6}$ in the region close to the (1,0) to 
(0,1) transition. In (c) and (d),
the gates have been slightly adjusted relative to (b) to allow
simultaneous transport and sensing. In (b) and (d), identical colour
scales are used. (Figure courtesy of C M Marcus.) Reprinted with permission from Petta J R,
Johnson A C, Marcus C M, Hanson M P and Gossard A C 2004 
{\it Phys. Rev. Lett.} \textbf{93} 186802. \copyright 2004 American Physical Society}\vspace{-0.5 cm}
\end{figure}

As already mentioned in section \ref{sec:Loss-DiVincenzo}, quantum dots can be created by electrical
gating of a 2DEG via lithographically
defined gate electrodes (see figures \ref{fig:twoqubitexchange}, 
\ref{fig:qdarray}, \ref{Fig:doubledot}, and \ref{fig: marcusdoubledot}).
Applying a negative voltage to the gates depletes
the 2DEG underneath them, such that quantum dots are formed in the
regions surrounded by the gates.
Electrically gated dots are typically
characterized by an electron level spacing $\delta\epsilon\approx0.1\dots2\:\mathrm{meV}$,
a charging energy $\delta\epsilon_{\mathrm{c}}\approx1\dots2\:\mathrm{meV}$,
and a dot diameter $l\approx10\dots1000\:\mathrm{nm}$ \cite{kouwenhoven:1997a,vanderwiel:2003a}.
Typical materials for such dots include GaAs, InSb, and Si. Control of the 
coupling of electrically gated GaAs quantum dots has been 
demonstrated and investigated in-depth in transport experiments 
\cite{vanderwiel:2003a,waugh:1995a,livermore:1996a}.

\begin{figure}[t]
\begin{center}
\includegraphics[%
  clip,
  width=14cm,
  keepaspectratio]{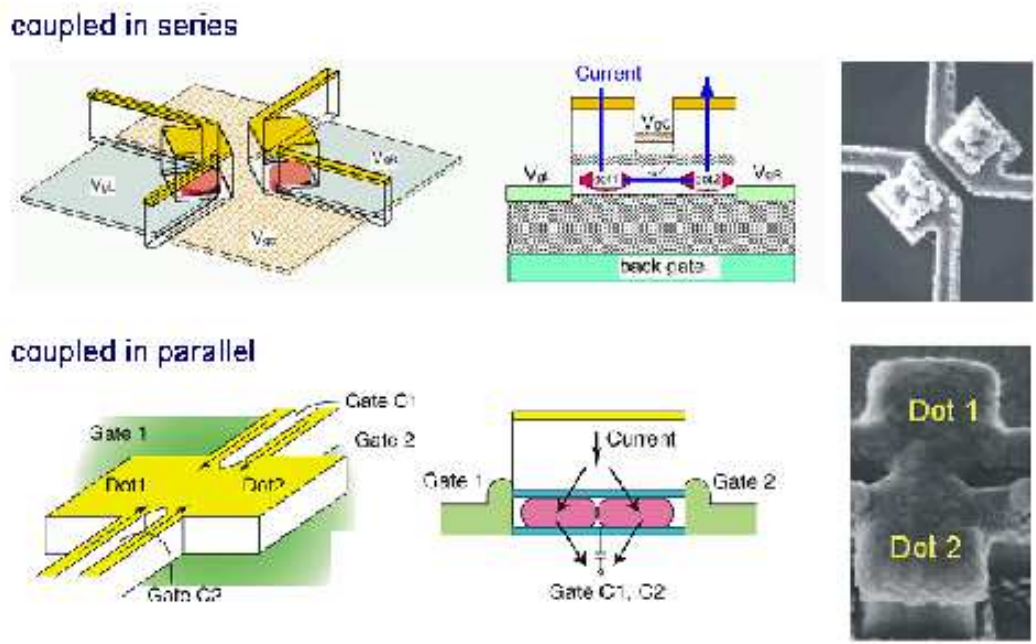}
\end{center}

\caption{\label{Fig: coupledqds} Different designs for etched structures
of coupled quantum dots \cite{hatano:2004a,kodera:2004a}. The upper three figures show two quantum dots that
can be probed by an electric current flowing through them in series,
whereas the lower three pictures show two dots that are coupled in parallel
for a transport experiment. The two rightmost figures are SEM micrographs.
(Figure courtesy of W G van der Wiel.) Reprinted from Kodera T, van der Wiel W
G, Ono K, Sasaki S, Fujisawa T and Tarucha S 2004 {\it Physica} E \textbf{22}
518. \copyright 2004, with permission from Elsevier.}
\end{figure}

As an alternative to electrical gating, etching techniques \cite{jacak:1998a}
can also be applied to achieve lateral confinement in the plane of a 2DEG.
For example, Tarucha {\it et al.} \cite{tarucha:1996a} have produced gated
vertical quantum dots by etching a pillar structure which contained
a double-barrier heterostructure with an InGaAs quantum well as the
2DEG. Figures \ref{Fig: coupledqds} and \ref{Fig: localESR}
show structures containing dots of this type.

Quantum dots also form ``naturally'' at monolayer steps at the interface 
of, e.g., thin GaAs/AlGaAs quantum wells. Usually,  molecular beam 
epitaxy (MBE) is used for the growth of such systems. If the MBE growth process
is performed without interruption, such steps occur at random positions as natural 
fluctuations of the quantum well width. Quantum dots of this type possess 
excellent optical 
properties, including very sharp optical linewidths. This has allowed 
the coherent control of optically excited states in experiments
\cite{bonadeo:1998a,gammon:2001a} and has recently culminated in the 
implementation of a {\sc crot} gate for qubits which are defined by the 
presence or absence of an exciton in the quantum dot \cite{li:2003a}.
  
\begin{figure}[t]
\begin{center}
\includegraphics[%
  clip,
  width=14cm,
  keepaspectratio]{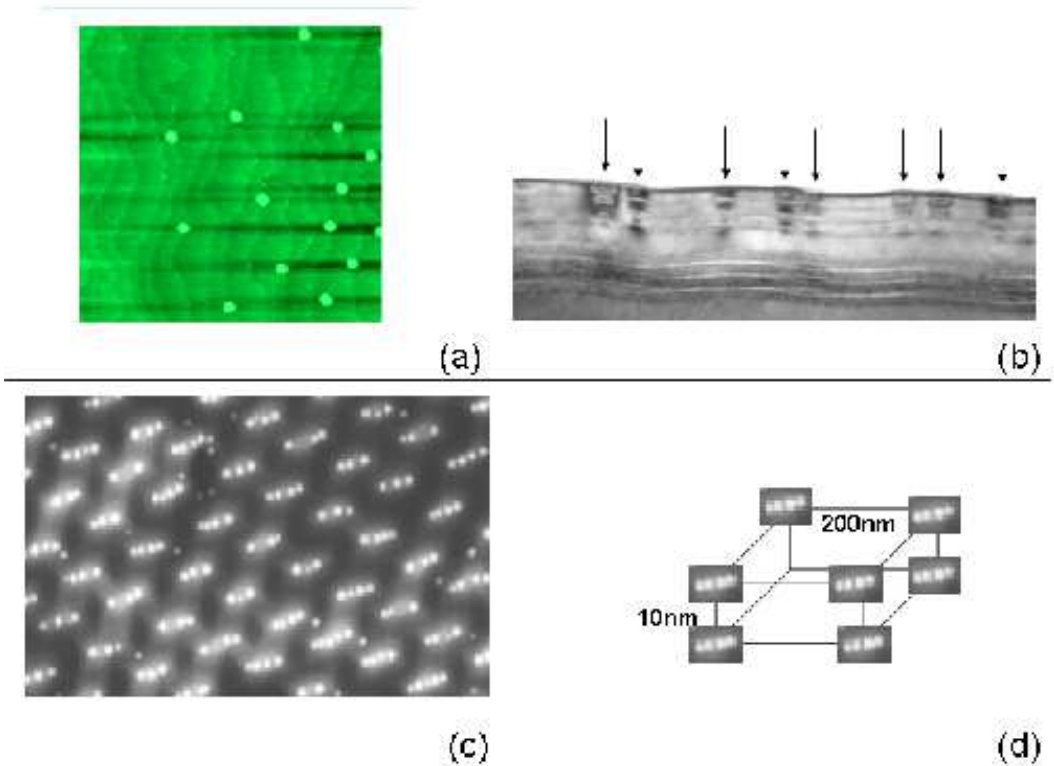}
\end{center}

\caption{\label{Fig: saqds} Self-assembled InAs quantum dot structures. (a)
AFM picture of dots grown at random locations. (b) Transmission electron
microscope (TEM) cross-section of vertically stacked dots (indicated by 
arrows), ordered
along the growth axis. (c) AFM picture of laterally ordered dots. 
This image was generated after
prepatterning of the substrate \cite{lee:2000a}. (d) Sketch of a three-dimensional
lattice of dots that could be obtained by combining the growth methods
of (b) and (c). (Figures courtesy of P M Petroff.)}
\end{figure}

Further, quantum dot structures can be grown by self-assembly, e.g., using
the Stranski-Krastanov growth technique. In this technique, self-assembled dot
islands form spontaneously during epitaxial growth due to a lattice
mismatch between the dot and the substrate material \cite{leonard:1993a}.
Typical sets of dot/substrate materials are InAs/GaAs, Ge/Si(100),
GaN/AlN, InP/GaInP, and CdSe/ZnSe \cite{eberl:2001a}. The electron
level spacing of this type of dot is typically $\delta\epsilon\approx30\dots50\:\mathrm{meV}$
with a charging energy $\delta\epsilon_{\mathrm{c}}\approx20\:\mathrm{meV}$,
a diameter $l\approx10\dots50\:\mathrm{nm}$, and a height $d\approx2\dots10\:\mathrm{nm}$
of the dot \cite{fricke:1996a}. Small self-assembled dots typically
have a pyramidal shape with four facets, whereas larger dots (containing,
e.g., 7 monolayers of InAs) form multi-faceted domes \cite{eberl:2001a}.
If pyramidal self-assembled dots are covered with a thin layer of
the substrate material (called the capping layer), the capped dots
 take-on an elliptical (or rarely, even a circular) shape.
Additionally, these dots exert strain on the capping layer. If quantum
dots are grown on the capping layer, they tend to grow on the strain
field on top of the capped dots rather than at random positions. 
This enables the growth of vertically
coupled quantum dots, where the thin capping layer acts as a barrier
between the two dots (see figure \ref{Fig: saqds} (b)). A typical
difficulty related to Stranski-Krastanov self-assembled dots is the
intrinsic randomness of the growth process, as shown in figure \ref{Fig: saqds}
(a). Yet, prepatterning of the substrate has been shown to be a way
to achieve a well-defined growth position of the first dot layer 
\cite{lee:2000a}
(see figure \ref{Fig: saqds} (c)), paving the way to site-controlled
arrays of single or coupled dots \cite{heidemeyer:2003a}. Cleaved-edge 
overgrowth is an alternative technique enabling atomically precise
control of the growth site of single and coupled dots \cite{schedelbeck:1997a}.
Colloidal chemistry is yet another promising approach to assemble quantum
dots with well-controlled size and shape \cite{alivisatos:1996a}.
Recently, colloidal CdSe dots have been coupled via molecular bridges
\cite{ouyang:2003a}. The inter-dot coupling in these experiments 
mediated coherent spin transfer
between the dots, which has subsequently been modelled theoretically
\cite{meier:2004a}. %

\subsection{Charge and spin control in quantum dots\label{sub:Charge-and-Spin}}

Precise control over the number of confined electrons has been demonstrated
several years ago in InGaAs self-assembled dots \cite{drexler:1994a},
in gated vertical quantum dots \cite{tarucha:1996a}, in quantum rings
\cite{warburton:2000a}, and also in electrostatically defined single
\cite{ciorga:2000a} and double \cite{elzerman:2003a,pioro:2003a,petta:2004a} dots in GaAs.
The single-electron states of quantum dots in the low-energy range
have been shown to be in agreement with a shell model. Because the
quantum dot confinement is much stronger along the growth direction than
perpendicular to it (for dots defined in a 2DEG as well as for self-assembled
dots), the dot potential is effectively two-dimensional. The low-lying
confined electron states can be well-approximated by the states of
a two-dimensional harmonic oscillator \cite{tarucha:1996a}. Thus,
the single-particle ground state has $s$ symmetry and the first excited
shell has $p$ symmetry. If an external magnetic field is applied
perpendicular to the quantum dot plane, new harmonic oscillator states
(Fock-Darwin states) are the exact eigenfunctions \cite{fockdarwin:19281930a},
with a frequency that increases with the magnetic field. Recently,
Raymond {\it et al.} \cite{raymond:2004a} have observed the Fock-Darwin
spectrum also for excitons (electron-hole pairs, rather than electrons 
alone) in quantum dots. 

\begin{figure}[t]
\begin{center}
\includegraphics[%
  clip,
  width=10cm,
  keepaspectratio]{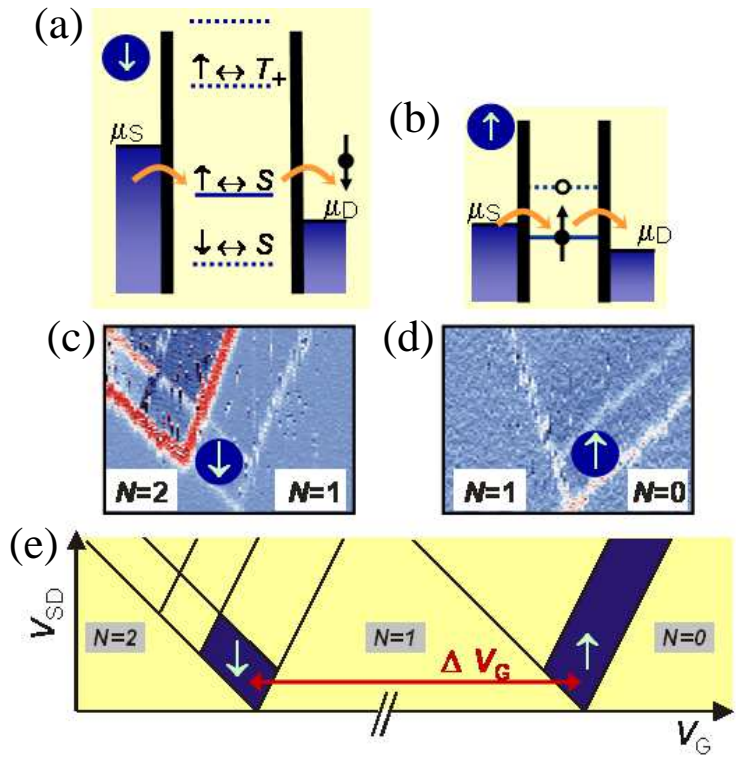}
\end{center}

\caption{Quantum dot spin filter \cite{hanson:2003a,elzerman:2004a}
(see sections \ref{sub:Spin-Relaxation} and \ref{sub:Spin-Initialization-and}). 
A static magnetic field splits the spin states of the quantum dot due 
to the Zeeman interaction.
For suitable gate voltages applied to the dot, the level configurations 
shown in (a) and (b) can be observed. In these cases, the 
transport through the dot is spin-dependent for sequential tunnelling. 
(a) Only electrons in the state $|\!\downarrow\rangle$ are 
transported through the dot. They form an 
intermediate singlet state $|S\rangle$ with an excess electron 
(being in the excited Zeeman level) on the dot. 
Tunnelling of electrons with spin 
$|\!\uparrow\rangle$ through the dot is energetically not possible 
because there is no intermediate two-electron state available
with an energy between the chemical potentials $\mu_{\mathrm{S}}$ and 
$\mu_{\mathrm{D}}$ of 
the source and the drain, respectively.
(b) Only the spin ground state,  $|\!\uparrow\rangle$, can
pass through the (empty) dot. 
In (c) and (d),  the measured differential conductance $dI/dV_{\mathrm{SD}}$ 
is shown for the cases (a) and (b), respectively, 
with tunnelling current $I$ and source-drain voltage 
$V_{\mathrm{SD}}$. In (e), we show a scheme of the theoretically predicted 
$dI/dV_{\mathrm{SD}}$ (which agrees well with (c) and (d)).
 \label{Fig: spinfilter} (Figure courtesy of L P Kouwenhoven.) Reprinted with
 permission from Engel H-A, Kouwenhoven L P, Loss D and Marcus C M 2004 {\it
 Quantum Inf. Process.} \textbf{3} 115. \copyright 2004 Springer-Verlag.}
\end{figure}

The degeneracy of the two spin states $|\!\uparrow\rangle$
and $|\!\downarrow\rangle$ is lifted in the presence of a magnetic
field due to the Zeeman interaction. This makes the two states energetically
distinguishable (see figure \ref{Fig: spinfilter}). The precise control of the occupation number of electrons
in single and double quantum dots has enabled experiments on single
spins in quantum dots, as we discuss in the following.

\subsection{Spin relaxation\label{sub:Spin-Relaxation}}

Recently, expectations for the stability of spin qubits in quantum
dots have grown considerably as progressively longer spin lifetimes
have been reported. A series of works on electron spin relaxation in
quantum dots started with Fujisawa {\it et al.} \cite{fujisawa:2002a}
who reported a triplet-to-singlet relaxation time of 
$\tau_{\mathrm{S-T}}=200\:\mu\mathrm{s}$ in vertical quantum dots. More recently, 
a lower bound on the singlet-triplet relaxation time has been measured in lateral 
dots, giving $\tau_{\mathrm{S-T}}\ge70\:\mu\mathrm{s}$ \cite{petta:2004b}.  Very quickly thereafter, 
a substantially longer relaxation time ($\tau_{\mathrm{S-T}}=(2.58\pm0.09)\:m\mathrm{s}$) 
was measured independently using a novel spin readout technique \cite{hanson:2004a}.  Several groups have 
since measured $T_{1}$ for {\it single} electron spins. For electrostatically-defined GaAs 
dots, Hanson {\it et al.} \cite{hanson:2003a} have
reported a lower bound $T_{1}\gtrsim 50\:\mu\mathrm{s}$ at a magnetic
field of $B=7.5\:\mathrm{T}$ which was subsequently topped by Elzerman
et al.\ \cite{elzerman:2004a}, with $T_{1}\approx(0.85\pm0.11)\:\mathrm{ms}$
at $B=8\:\mathrm{T}$. In these experiments, a two-level pulse technique
for the quantum dot gate voltage has been applied to inject an electron
into the dot and to extract it later. In a certain parameter range,
the Zeeman splitting of the two spin states is sufficient that tunnelling
into or out of the dot is not possible for one of the two spin states
\cite{recher:2000a,ciorga:2000a,fujisawa:2002a,hanson:2003a} 
(see also figure \ref{Fig: spinfilter}). This enables spin detection
via the detection of charge in the quantum dot, which has been realized
through an adjacent quantum point contact (QPC) 
\cite{field:1993a,elzerman:2003a,hanson:2003a,elzerman:2004a} 
(in a setup  similar to that 
shown in figure \ref{Fig:doubledot} (a) for a double quantum dot).
In these experiments, the QPC has been tuned via a gate voltage to
a conductance $G\approx e^{2}/h$, where the modulation of the current
$I_{\mathrm{QPC}}$ through the QPC has maximum sensitivity to changes in the
electrostatic environment, including the number of charges in the
quantum dot. Recently, Kroutvar {\it et al.} \cite{kroutvar:2004a} established
a lower bound $T_{1}\gtrsim 20\:\mathrm{ms}$ at $T=1\:\mathrm{K}$
and $B=4\:\mathrm{T}$ for In(Ga)As self-assembled dots. 
In this experiment, an optical
charge storage device has been excited with circularly polarized laser
excitation. The larger level spacing of self-assembled 
dots (compared to gated GaAs dots) is responsible for the longer $T_1$-time
 seen in this experiment which is limited by spin-orbit coupling
(see also section \ref{subsub:Spin-orbit}).

\subsection{Spin decoherence\label{sub:Spin-Decoherence}}
The spin coherence of electrons localized at impurity centres has 
been investigated deeply for the last few decades in ensemble measurements.
Many of these experiments have investigated the spin dephasing of electrons 
bound by the Coulomb interaction to a donor in silicon (for example,
phosphorus, antimony, or arsenic). The wavefunction of such donor-bound 
electrons is quite similar to the wavefunction of electrons bound in a 
quantum dot.
Several of these experiments have 
demonstrated rather long electron spin decoherence times,
which is mainly due to the confinement of the electrons in all three spatial 
dimensions (leading to a $\delta$-peaked density of states).
The electron nuclear double resonance (ENDOR) method has been applied to 
map-out the wave function of the bound electron \cite{feher:1959a}.
Hahn-echo measurements have shown that $T_2 \approx 10^{-4}\:\mathrm{s}$ 
for donor electron spins in phosphorus-doped silicon (Si:P) 
\cite{gordon:1958a}. Recent spin-echo measurements of 
isotopically purified $^{28}$Si:P  have shown that $T_2 = 62 \: \mathrm{ms}$ 
\cite{tyryshkin:2003a}. This very long $T_2$-time is possible in such 
systems since $^{28}$Si has nuclear spin $I=0$, drastically reducing the 
hyperfine interaction.
In contrast, spin-echo measurements of electron spins bound to 
$^{29}$Si:P donors in isotopically purified $^{29}$Si have shown 
a much shorter envelope decay time (essentially $T_2$) on the order of
$T_M \approx 10^{-5}\:\mathrm{s}$ \cite{itoh:2004a}.

To our knowledge, there are only very few results published on measurements
of the $T_{2}$ time of single electron spins in {\it quantum dots}. Still,
optical experiments probing the decoherence time of \emph{exciton}
spins may provide a lower bound for the $T_{2}$-time of single electrons. 
Gupta {\it et al.} \cite{gupta:1999a} have measured a lower bound for the ensemble
dephasing time of $T_{2}^{*}\approx3\:\mathrm{ns}$ for CdSe dots
using femtosecond-resolved Faraday rotation. In this experiment, different
decay time scales have been observed for the spin precession, showing a 
more complicated dynamics than expected.
Recent $g$-factor calculations for electrons and holes in CdSe dots, 
based on time-dependent 
empirical tight-binding theory, addressed this issue \cite{chen:2004a}. 
A strongly anisotropic $g$-factor, with $g_x\approx g_y > g_z$ for all 
dot sizes (where $z$ denotes the $c$-axis of the wurtzite crystal)
has been obtained for the electron. The range of 
$g$-factors (for the corresponding dot sizes) is in agreement with the 
experimentally \cite{gupta:1999a} extracted pairs of $g$-factors,
providing a first step in the understanding of the observed nontrivial 
dynamics of electron and hole spins in quantum dots. 
Measuring the Hanle effect
in an ensemble of InAs self-assembled dots, Epstein {\it et al.} 
\cite{epstein:2001a}
obtained $g_{e}T_{2}^{*}\approx210\:\mathrm{ps}$ at $T=6\:\mathrm{K}$,
where $g_{e}$ is the electron $g$-factor. In contrast to the 
single-spin decoherence time $T_2$, the ensemble
dephasing time $T_2^*$ might be reduced from $T_2^*=T_2$ by dephasing 
among the spins of 
the measured ensemble.
Further, the electron-hole exchange
interaction couples electron and  hole spins in experiments that involve excitons. It can be 
assumed that this coupling further influenced 
the decay of the observed luminescence  polarization. 
It might thus be possible that the coherence of single electron spins is larger
than the values obtained from these experiments. 
In fact,
recent Hanle measurements on individual quantum dots \cite{bracker:2004a} 
have indicated an
electron decoherence time $T_2\approx 16\:\mathrm{ns}$. Yet, this result 
may have slightly exceeded the expected value $T_2\approx \hbar\sqrt{N}/A$,  
 discussed in section \ref{sec:hyperfine}. The quantum dots in this experiment were 
defined by monolayer-high steps at the interfaces of a
3 nm thick GaAs/AlGaAs quantum well.

In section
\ref{sub:Goals}, we discuss further proposals to measure the $T_{2}$-time 
of a single electron spin in a quantum dot. Given the measured $T_{1}$
values in the millisecond range and measured $T_2$ times that are far 
smaller, it can be expected that nuclear spins
are typically the dominant source of decoherence for electron spins in quantum
dots.

\subsection{Spin initialization \label{sub:Spin-Initialization-and}}

To initialize the spin qubits, a strong polarization can be achieved
by applying a strong magnetic field $B$, such that the Zeeman splitting
is larger than the thermal energy, as already mentioned in section 
\ref{sec:Loss-DiVincenzo}. 
Further, electrons with
parallel spins can also be injected via spin-polarized currents. The
injection of spins from ferromagnetic semiconductors into normal 
semiconductors have
been reported with polarizations up to 90\% \cite{fiederling:1999a,ohno:1999a}.
Initialization as well as detection of a single spin can also be achieved 
using a spin filter (see section \ref{sec:spindetection}) 
or by optical schemes (see section \ref{sec:optical initialization}).

\subsection{Single-spin detection \label{sec:spindetection}}
A central question for the readout of a single spin is the reliability
of the experimental result. We briefly address this issue here. Errors
during the measurement process can be eliminated statistically by
performing an experiment $n$ times identically. This procedure is called
$n$-shot readout. There is a probability $p$ that the experimental
readout procedure of a certain quantum mechanical state yields the correct
result, and a probability $1-p$ that it does not. In this way,
one can define the probabilities $p_{\uparrow}$ and $p_{\downarrow}$
for the measurement successes of the states of a spin-$1/2$. 
Including the possibility of an error,  
the measurement of the state of a spin-1/2  
 is described by a measurement of the observables
\begin{eqnarray}
A_\uparrow & = & p_\uparrow |\!\uparrow\rangle\langle\uparrow \!|+
(1-p_\downarrow)|\!\downarrow\rangle\langle\downarrow \!|,\\
A_\downarrow & = & p_\downarrow |\!\downarrow\rangle\langle\downarrow \!|+
(1-p_\uparrow)|\!\uparrow\rangle\langle\uparrow \!|,
\end{eqnarray}
where $A_\uparrow$ is the observable leading to the experimental result 
``spin up'', whereas
$A_\downarrow$ leads to the result ``spin down''.
To achieve a reliable measurement up
to a significance level (``infidelity'') $\alpha$, a statistical
analysis of the readout process \cite{engel:2003a} yields the result that the
number $n$ of required measurements has a lower bound
\begin{equation}
n > z_{1-\alpha}^{2}\left(\frac{1}{\eta}-1\right),\label{eq:nshotreadout}
\end{equation}
 where $z_{1-\alpha}$ is the quantile (critical value) of the standard
normal distribution function, $\Phi(z_{1-\alpha})=1-\alpha=(1/2)[1+\mathrm{erf}(z_{1-\alpha}/\sqrt{2})]$,
and 
\begin{equation}
\eta =\left(\sqrt{p_{\uparrow}p_{\downarrow}}-\sqrt{(1-p_{\uparrow})(1-p_{\downarrow}})\right)^{2}\label{eq:efficiency}
\end{equation}
 can be interpreted as a measurement efficiency with $\eta\in[0,1]$.
For example, if $p_{\uparrow}=1-p_{\downarrow}$, it is not possible to 
distinguish  between the two spin states and $\eta =0$. 
In contrast, for $p_{\uparrow}=p_{\downarrow}=1$,
the measurement is perfectly reliable and $\eta =1$. The case $n=1$
(which is realized, e.g., in the latter example) is called single-shot
readout in the following. For a set of $k$ qubits, the probability
for a reliable measurement is given by $1-\beta=(1-\alpha)^{k}$, where 
$\beta$ is the infidelity of the $k$-qubit readout.
However, the number $n$ of required measurements only grows with
$k$ according to $n\geq 2\left(1/\eta -1\right)\log k /\beta$
\cite{engel:2004a}. The dependence of $n$ on $k$ is therefore weaker than
what might be naively expected.

\begin{figure}[t]
\begin{center}
\includegraphics[%
  clip,
  width=14cm,
  keepaspectratio]{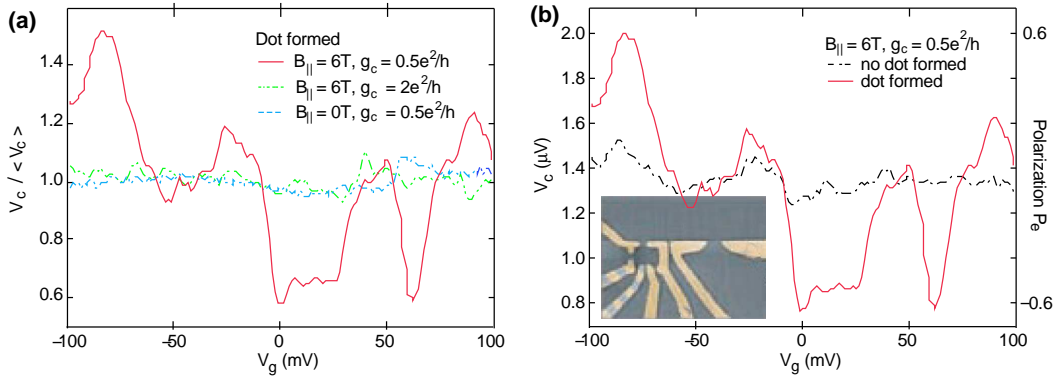}
\end{center}

\caption{\label{Fig: spinfilter2} Experimental demonstration of a spin 
filter \cite{folk:2003a}. The inset shows a micrograph of the 
structure used, where the quantum dot on the left-hand 
side is the polarizer and the QPC on the right-hand side is the analyzer.
The figures (a) and (b) show the focusing peak height as a function of the 
quantum dot gate voltage $V_{\mathrm{g}}$.
See section \ref{sec:spindetection} for a description of the 
experiment.
(a) Large fluctuations of the focusing peak height 
are measured at in-plane magnetic field $B_\| = 6\:\mathrm{T}$ if 
the collector is spin-selective (solid line). These  
fluctuations  
are greatly reduced if the conductance of the QPC used as 
the collector is tuned out of the spin-selective regime \cite{potok:2002a} 
(dotted line) or for zero in-plane magnetic field, $B_\| =0$ 
(dashed line).
(b) The spin-filter effect is detectable at 
$B_\| = 6\:\mathrm{T}$ with spin-selective collector when the emitter is 
a quantum dot (solid line) and vanishes if the quantum dot is 
transformed into a QPC on the $2e^2/h$ plateau (dot-dashed line).
(Figure courtesy of C M Marcus.) Reprinted with permission from Engel H-A,
Kouwenhoven L P, Loss D and Marcus C M 2004 {\it Quantum
  Inf. Process.} \textbf{3} 115. \copyright 2004 Springer-Verlag.}
\end{figure}

The magnetic moment of a single spin-$1/2$ is very small (on
the order of $\mu_{\mathrm{B}}=9.2741\cdot10^{-24}\:\mathrm{J/T}$)
and thus difficult to detect directly. Nevertheless,
Rugar {\it et al.} \cite{rugar:2004a} have recently detected a single 
spin in silicon dioxide
using MRFM, as already mentioned in section \ref{sec:silicon}.  MRFM enables the
direct observation of an oscillating spin up to $100\:\mathrm{nm}$ 
below the surface with nanometre resolution. Still, the sensitivity is
currently not yet sufficient to detect whether a spin is 
originally in the state 
$|\!\uparrow\rangle$ or in the state 
$|\!\downarrow\rangle$.
Many other proposals to detect spin
states are based on the transfer of information stored in the
spin degree of freedom to an orbital degree of freedom (``spin-charge 
conversion'') 
\cite{loss:1998a,hanson:2003a,elzerman:2004a,recher:2000a,folk:2003a,potok:2003a,cortez:2002a,shabaev:2004a,friesen:2004a,engel:2001a,engel:2002a}.
Initialization and readout of spin states in quantum dots can be achieved,
e.g., using a spin filter. This is a device that only transmits electrons
with one particular spin polarization, while the opposite spin 
polarization is blocked. Recher
{\it et al.} \cite{recher:2000a} have proposed a spin-filter implementation
consisting of a quantum dot in the Coulomb blockade regime,
weakly coupled to two current leads. In a
static magnetic field, the direction of the transmitted spin 
can be changed by tuning the gate voltage applied to the dot (see figures 
\ref{Fig: spinfilter} and \ref{Fig: spinfilter2}). Experimental
demonstrations of a spin filter have been achieved by 
Folk {\it et al.} \cite{folk:2003a},
Potok {\it et al.} \cite{potok:2003a}, Hanson {\it et al.} \cite{hanson:2003a},
and Elzerman {\it et al.}  \cite{elzerman:2004a}. The first two of these
implementations have demonstrated the spin-filtering effect with a GaAs
quantum dot in the open \cite{folk:2003a} and in the Coulomb-blockade
regime \cite{potok:2003a} in a polarizer-analyzer geometry (see also figure 
\ref{Fig: spinfilter2}). 
In the polarizer-analyzer geometry (see inset of figure \ref{Fig: spinfilter2}),
the spin-selective analyzer was provided by a QPC with conductance tuned to 
less than $e^2/h$ \cite{potok:2002a}. A small perpendicular magnetic field 
$B_{\perp}$ coupled the polarizer (i.e., the quantum dot structure to the left) 
and the analyzer (the QPC to the right) by 
transverse focusing.
A transverse magnetic field $B_{\|}$ was applied, leading to a different Fermi 
wavelength of spin-up and spin-down electrons. By tuning the gate voltage 
of the dot, the transmission of one or the other spin was suppressed due to 
destructive interference of the coherent transport paths. 
With a constant current flowing between emitter (i.e., the dot) and 
collector (i.e., the QPC), peaks were 
observed in the voltage $V_c$  between collector and base whenever the 
distance between emitter and collector was an integer multiple of the 
cyclotron diameter of the transported electrons. In an in-plane magnetic 
field $B_{\|}$, the height of these peaks in $V_c$ (which are called 
``focusing peaks'') reflected the degree of spin polarization in the current
if the QPC was in the spin-selective regime.
The experiments 
by Hanson {\it et al.} \cite{hanson:2003a} and 
Elzerman {\it et al.} \cite{elzerman:2004a}
 have already been
described in section \ref{sub:Spin-Relaxation}. 
Elzerman {\it et al.} \cite{elzerman:2004a}
demonstrated \emph{single}-shot readout of a single electron spin
in a quantum dot. A single-spin measurement of this type required
a time $\approx0.11\:\mathrm{ms}$ and the total fidelity of the spin
readout was estimated to be $65\%$. %

\subsection{Optical interaction and optical readout of spins \label{sec:optical interaction}}
In this section, we first sketch some basics of optical transitions in quantum dots and then focus on the optical detection of spin states.
 The currently very active field of ultrafast laser technology suggests
that  single spin states can be optically detected and manipulated within
very short times (picoseconds or even femtoseconds), several orders
of magnitude faster than in schemes based on the transport of electric
charge. 

Via the absorption of a photon, 
an electron in a confined
valence-band state can be excited to a confined conduction-band state. 
For such inter-band transitions, optical selection rules apply and establish 
conditions on the quantum numbers of the optically coupled states.
Provided the spin-orbit interaction is nearly isotropic 
($H_{\mathrm{so}} \approx \lambda \mathbf{L} \cdot \mathbf{S}$, see 
also the discussion in section \ref{subsub:Spin-orbit}), then it is a 
good approximation that the total angular 
momentum squared, $\mathbf{J}^2=(\mathbf{L}+\mathbf{S})^2$, provides a 
good quantum number
in semiconductors.   
Photons with circular polarization $\sigma^{\pm}$ carry an angular 
momentum with projection $\pm 1$ (in units of $\hbar$) along their propagation direction. 
For optical interactions, the total angular momentum is conserved, linking 
the spin of electrons and the polarization of photons.  
For a two-dimensional  quantum dot with circular confinement,
the $z$ component $J_z$ of $\mathbf{J}$ 
is a good quantum number 
 (in contrast, an anisotropic 
shape in the plane induces mixing of angular momentum eigenstates). 
When $J_z$ is a good quantum number in GaAs or InAs dots, the
energetically lowest optical excitation at zero magnetic field typically
includes two degenerate valence band states with total angular momentum
projections $J_{z}=\pm3/2$, which are also called heavy-hole (hh)
states. A circularly polarized photon
that is irradiated along the quantization axis $z$ of $\mathbf{J}$ 
 can excite one of the hh states to one of the 
conduction-band states with spin $+1/2$ or $-1/2$ \cite{dyakonov:1984a}. 
For a given circular polarization, only one combination 
of these states satisfies the selection rules. 
This leads to a direct correspondence between the circular polarization
of the photon and the spin of the optically excited electron. 
Taking advantage of this for the readout of spin states, light-emitting 
diodes (``spin-LEDs'') 
have been fabricated \cite{fiederling:1999a,ohno:1999a}, where the polarization
of the emitted photons indicates the spin polarization of the electrons 
(or holes) injected into the spin-LED. 
A further step in nanoscale photonic and electronic 
technology has been taken recently by the growth of semiconductor nanowire superlattices 
\cite{wu:2002a,bjork:2002a,gudiksen:2002a}.
By modulating the reactants during catalytic growth of a nanowire,
the nanowire finally consists of segments of different materials, e.g.,
Si and SiGe \cite{wu:2002a}, InAs and InP \cite{bjork:2002a}, or 
GaAs and GaP \cite{gudiksen:2002a}.
By alternating the two different materials, a superlattice can be formed.
The combination of $n$- and $p$-type semiconductors, e.g., 
$n$-Si and $p$-Si or $n$-InP and $p$-InP \cite{gudiksen:2002a}, 
enables the bottom-up assembly of nanoscale (spin-)LEDs.  

\subsection{Negatively charged excitons in quantum dots \label{sec:charged exciton}}
\begin{figure}[t]
\begin{center}
\includegraphics[%
  clip,
  width=8.6cm,
  keepaspectratio]{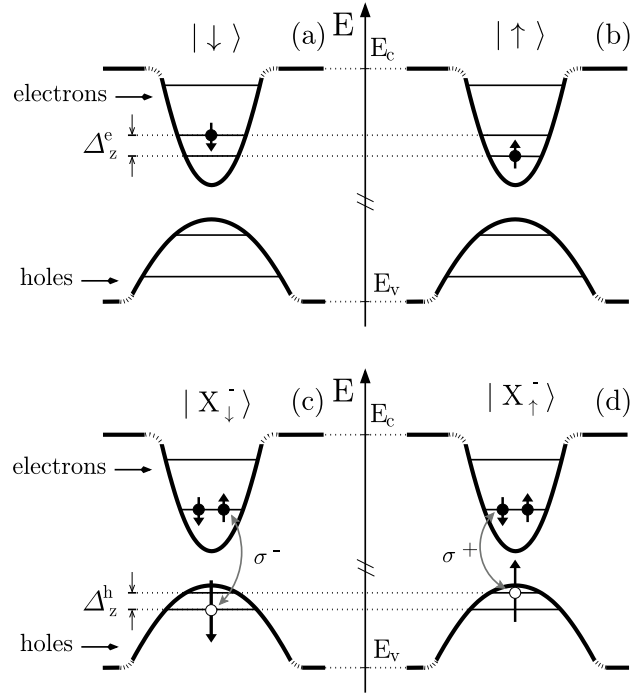}
\end{center}

\caption{\label{Fig: trions} The two Zeeman states $|\!\downarrow \rangle$ 
and $|\!\uparrow \rangle$ of a single 
electron in the dot are shown in (a) and (b), whereas (c) and
(d) show the two Zeeman states $|X_{\uparrow}^{-}\rangle$ and 
$|X_{\downarrow}^{-}\rangle$ of a negatively charged exciton in
the orbital ground state \cite{gywat:2004a,gywat:2004b}. As discussed in
the text, the two electrons in (c) and (d) form a spin singlet. The grey arrows
indicate which electron-hole pair is coupled by a $\sigma^{\pm}$
circularly polarized transition. In the presence of a static magnetic
field along the $z$ direction, the Zeeman splitting of the electron
spin states is $\Delta_{z}^{e}$ and the Zeeman splitting of the charged
exciton equals the hole Zeeman splitting $\Delta_{z}^{h}$. Here, we assume equal signs for the $g$-factors of electrons and holes.}
\end{figure}
Several
methods have been developed to optically probe and manipulate states
of \emph{single} quantum dots \cite{grundmann:1995a,dotspectroscopy}. 
Optical schemes have further been 
proposed to achieve 
initialization of electron spins
(see section \ref{sec:optical initialization}), for the detection
of the $T_2$-time of electron spins (see section \ref{sub:T2measurements}), 
for single-qubit gates (see section \ref{sub:Single-Qubit-Rotations}), 
and for two-qubit gates (see section \ref{sub:Two-Qubit-Gates}).
In these schemes and also in many other schemes exploiting the spin 
states of an electron,
a quantum dot initially contains a single excess electron.
Optical excitation of such a state
creates a negatively charged exciton (sometimes also called
``trion'') in the dot, i.e., a compound of two conduction-band
electrons and one valence-band hole (see figure \ref{Fig: trions}).
If the quantum dot is in the so-called strong confinement regime,
the (single-particle) confinement energies are much larger than the
Coulomb interaction energies of the carriers in the dot. This criterion
is typically satisfied for small self-assembled dots and colloidal dots.
The two electrons then occupy the lowest single-particle level of
the dot and form a spin singlet. 
Note that the
excess electron initially occupies one of the available spin states.
Due to the Pauli principle, the absorption of a circularly polarized 
photon (as described in section \ref{sec:optical interaction}) is only possible if the corresponding 
electron spin state is not already occupied. Figure 
\ref{Fig: trions}  shows that a  $\sigma^-$-polarized photon
can only be absorbed if the spin of the excess electron is in the state 
$|\!\downarrow\rangle$, whereas  a $\sigma^+$-polarized photon can only be 
absorbed for $|\!\uparrow\rangle$.
In the photoluminescence
spectrum, the lines belonging to these two transitions
 coincide for zero magnetic field and  split
for non-zero magnetic fields. 
If a circularly polarized photon with an energy that matches the corresponding 
transition energy is absorbed, the initial spin state of the excess 
electron is identified. 
This experiment has recently been 
performed
with a single InGaAs/GaAs dot by H\"ogele {\it et al.} \cite{hoegele04a}
using high-resolution laser absorption spectroscopy. Equivalently,
the photoluminescence (which is only emitted after a successful photon 
absorption)
could be detected instead of the absorption. One can also apply an electric
field to the dot such that an electron and a hole tunnel out of the
dot after a photon has been absorbed. Instead of the photoluminescence,
the resulting electric current (the so-called photocurrent) can then
be detected \cite{baier:2001a}. For a discussion of the limits of such 
spin-dependent optical schemes due to the mixing of valence-band states,
see section \ref{sec:optical initialization}.

\subsection{Optical initialization of spin qubits \label{sec:optical initialization}}

The spin of an excess electron in a quantum dot can be polarized for 
initialization by using
optical pumping methods \cite{cortez:2002a,shabaev:2004a,gywat:2004a}.
As discussed in section \ref{sec:charged exciton}, circularly polarized laser excitation can be used
to optically address exclusively one of the two spin states 
$|\!\uparrow\rangle$
or $|\!\downarrow\rangle$. For initialization of a spin, the optical
excitation can also be tuned to higher-lying continuum states 
\cite{cortez:2002a}.
Alternatively, applying circularly polarized optical $\pi$-pulses
in the presence of a static (or pulsed) transverse magnetic field also
increases the electron spin polarization \cite{shabaev:2004a}. 
In this scheme,
the transverse magnetic field has a negligible effect on the charged
exciton states because the in-plane $g$-factor of the hole is typically
zero in first order (the response of the charged exciton
 to an external magnetic field is, in this case, determined by the
hole spin since the two electrons form a singlet). A circularly polarized 
photon can now be absorbed for only, say, $|\! \downarrow\rangle $. After the absorption of a photon, 
the precession of the spin is locked until, after recombination, the 
initial spin state $|\! \downarrow\rangle $ is restored.
The other electron spin state, $|\! \uparrow\rangle $, blocks the photon 
absorption and is therefore
 rotated by the transverse magnetic
field without interruption. By choosing
suitable pulse repetition rates and magnetic field strengths, the
spin is polarized in the state $|\! \downarrow\rangle $. 
Yet another
way to achieve electron spin polarization is to apply a magnetic field
parallel to the laser beam and choose the circular polarization of
the laser such that the hole contained in the charged exciton is in its excited
Zeeman level, as shown in figure \ref{Fig: trions}
(c). Hole spin relaxation within the charged exciton (which occurs at elevated
temperatures at even larger rates than those for optical 
recombination~\cite{flissikowski:2003a}) and subsequent recombination
leads then to an increased polarization of the electron spin in the state
$|\! \uparrow\rangle $ that does not allow photon absorption \cite{gywat:2004a} (in contrast to 
the scheme mentioned above).  
To benefit from Pauli blocking of the absorption in these schemes,
the bandwidth of the laser must be smaller than the splitting of
hh and, typically, light hole (lh) states (which have angular momentum 
$J_{z}=\pm1/2$) and also the energy difference to the state with 
one electron in the first excited level, forming a triplet state 
with the electron in the orbital ground state. For self-assembled 
dots, this hh-lh splitting is on the order of $10\:\mathrm{meV}$, and 
the energy difference to the mentioned state with an electron triplet
is approximately $40\:\mathrm{meV}$ \cite{cortez:2002a}.

In the context of optical transitions including  hh and lh states 
it has turned out that the geometry of the quantum dot can also impose a
limitation on the efficiency of spin-dependent optical processes.
As already mentioned in section \ref{sub:Structures-of-Single}, capped
self-assembled dots are elliptical, rather than circular in
shape. This anisotropy leads to a mixing especially of the valence
band states (since they are close in energy to each other). 
If the bandwidth of the circularly polarized laser is
larger than the Zeeman splitting of the electron states, the admixture
of, e.g., lh states 
with the hh
states (as in bulk semiconductors) increases the probability that a 
photon is absorbed even though
the spin is in the state where Pauli blocking should be effective
 \cite{dyakonov:1984a}.
However, if a circularly polarized laser with a bandwidth smaller
than the electron Zeeman splitting is applied in resonant optical
experiments \cite{hoegele04a}, the mixing of the hole states has no
effect on the absorption properties of the quantum dot because then, 
again, 
only one of the electron spin states allows for photon absorption.

\section{Future Goals\label{sub:Goals}}

In this section we discuss recent theoretical proposals to measure
the $T_{2}$ time of single electron spins in quantum dots and also
proposals for single-qubit rotations.

\subsection{Detection of single-electron spin decoherence\label{sub:T2measurements}}

After the recent successful measurements of the $T_{1}$-lifetime 
(and lower bounds for it) of single electron spins in quantum dots 
(see section \ref{sub:Spin-Relaxation}),
measurements of the decoherence time $T_{2}$ are due. To achieve
such an experiment, an initial coherent evolution of the electron
spin must be produced. This can be done, e.g., with electron
spin resonance (ESR) or by inducing spin precession in a transverse
magnetic field. The decay of the spin coherence can then be measured
\cite{bracker:2004a}.
Several proposals of this type have been made. Engel and Loss 
\cite{engel:2001a,engel:2002a} have
proposed a measurement of the sequential tunnelling current through a dot
containing a single electron spin in the presence of ESR excitation.
Sequential tunnelling, in general, describes a regime where charge 
transport only occurs via a sequence of first-order tunnelling processes.  In the regime when sequential tunnelling  is only possible via an intermediate
singlet state on the dot \cite{engel:2001a,engel:2002a}, the stationary
current $I$ is a Lorentzian as a function of the ESR detuning $\delta_{\mathrm{ESR}}=\omega_{\mathrm{ESR}}-g_{e}\mu_{\mathrm{B}}B$,
where $\omega_{\mathrm{ESR}}$ is the ESR frequency. The inverse of
the linewidth of $I(\delta_{\mathrm{ESR}})$ provides a lower bound
for the intrinsic $T_{2}$ time of a single electron spin. Further,
the coherent Rabi oscillations due to ESR pulses can be observed in
the time-averaged current $\bar{I}(t_{\mathrm{p}})$ as a function
of the ESR pulse length $t_{\mathrm{p}}$. Subsequently, 
Martin {\it et al.} \cite{martin:2003a} have
proposed the electrical detection of single-electron spin resonance
via a nearby field-effect transistor conduction channel. In contrast
to a transport measurement, Gywat {\it et al.} \cite{gywat:2004a,gywat:2004b}
have theoretically studied the optical detection of magnetic resonance 
(ODMR) to measure
the $T_{2}$-time of a single electron spin in a quantum dot. 
In this approach, the dot
 initially contains a single excess electron that is subject
to ESR excitation. Unlike a tunnelling experiment 
\cite{engel:2001a,engel:2002a,martin:2003a},
optical transitions are subject to selection rules and are not restricted
to the Coulomb blockade regime, e.g., if the excess electron is present
due to \emph{n}-doping and is not electrically injected. Further,
an ODMR experiment can be performed without connecting the dot to
current leads, which reduces decoherence. One can additionally benefit
from the high sensitivity of photodetectors. For a $\sigma ^-$-polarized 
laser with a sufficiently
low bandwidth the absorption of a photon is Pauli-blocked if the
 spin is in the state $|\! \uparrow\rangle$, as discussed in section 
\ref{sec:charged exciton} (see also figure \ref{Fig: trions}).
The laser frequency and polarization ($\sigma^-$) in the considered ODMR scheme
are adjusted such that in the case of successful photon absorption,
 a negatively charged exciton, as shown in figure \ref{Fig: trions}
(c), is created, where the two electrons form a singlet and the hole
is in the excited Zeeman level of the orbital ground state. From here,
there are two possible relaxation paths, either the direct optical
recombination, or a hole spin flip and an optical transition with
opposite circular polarization. This second relaxation channel is
responsible for an accumulation of population in the spin ground state
(exactly as discussed  in section \ref{sec:optical initialization}
for spin initialization) since the optical recombination rate is 
usually much
faster than the ESR Rabi frequency. For cw ESR and cw laser excitations,
the stationary photoluminescence \cite{gywat:2004a} or, alternatively,
the stationary photocurrent \cite{gywat:2004b} has been found to
be a Lorentzian as a function of the ESR detuning $\delta_{\mathrm{ESR}}$.
As in the detection of the ESR linewidth using sequential
tunnelling, the inverse linewidth of the photoluminescence or the photocurrent
provides a lower bound for $T_{2}$. Additional broadening due
to the optical transitions is greatly reduced for a hole spin flip rate
that is comparable to or larger than the optical recombination rate,
as well as for an optical Rabi frequency on the order of $1/T_{2}$
or smaller. Alternatively, pulsed laser excitation can be applied
in addition to an ESR excitation. This enables the detection of spin
Rabi oscillations as a function of the laser pulse repetition time
$\tau_{\mathrm{rep}}$. Because of hole spin flips, the electron spin
at the end of a laser pulse is polarized as mentioned above. 
During the ``off''-time of the laser, the spin is performing Rabi 
oscillations. When the subsequent laser 
pulse arrives,  the spin state $|\! \downarrow\rangle$ is read out. The 
time-averaged number of photons that are emitted per laser repetition 
period then directly displays the electron spin Rabi oscillations as a 
function of $\tau_{\mathrm{rep}}$.
Increasing
the length of the laser pulses to values longer than the exciton lifetime
iterates the optical pumping scheme and therefore enhances its efficiency.
This results in an improved visibility of the oscillations in the
photoluminescence or in the photocurrent. Using the same optical excitation
setup, electron spin precession can also be observed in the presence
of a transverse magnetic field \cite{gywat:2004b}.

\subsection{Single-qubit rotations \label{sub:Single-Qubit-Rotations}}

\begin{figure}[t]
\begin{center}
\includegraphics[%
  clip,
  width=8.6cm,
  keepaspectratio]{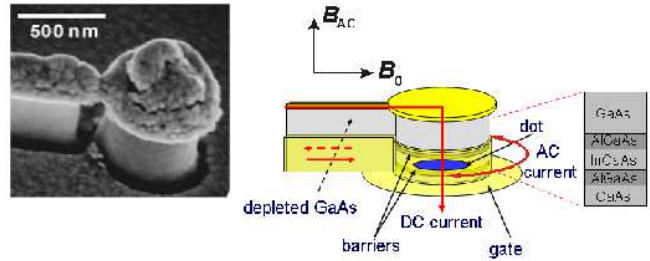}
\end{center}

\caption{\label{Fig: localESR} SEM picture and scheme of a structure to apply
a local rf magnetic field to a quantum dot. Such a structure might be used
as a prototype of a single qubit gate, or for the measurement of the
electron spin decoherence time. The indicated AC current (in the horizontal
direction) leads to an alternating magnetic field $B_{\mathrm{AC}}$.
In combination with a static magnetic field $B_{0}$, ESR can be induced
with an electron located in the dot. From the modulation of the DC
current (in the vertical direction) as a function of the frequency
of the AC current, the electron spin decoherence time can be measured
\cite{engel:2001a,engel:2002a}. (Figure courtesy of W G van der Wiel \cite{kodera:2005a}.) }
\end{figure}

A further important step towards the goal of quantum computation is the 
implementation of a single-qubit gate.
To achieve this for the Loss-DiVincenzo proposal, several possible 
strategies have been developed \cite{loss:1998a,burkard:1999a,awschalom:2002a}. 
The simplest way to rotate a spin is by applying a pulsed magnetic field.
In an array of quantum dots, such fields could be applied to single spins, 
e.g., by scanning-probe tips \cite{loss:1998a}.
Further, in the presence of
an rf magnetic field applied to an ensemble of electron spins, the tunability
and precise control of the individual Zeeman splittings is sufficient
to produce single spin rotations, as already mentioned in section 
\ref{sec:Loss-DiVincenzo}. When
the ESR resonance condition is matched, the spin rotates with maximum
amplitude, according to the well-known Rabi formula. Detuning of the
Zeeman splitting of an individual spin from the ESR resonance slows 
its precession frequency and the spin stops rotating entirely
when the detuning is larger than the ESR linewidth. Control of
the Zeeman splitting at the single-spin level is therefore another way to
perform single-spin rotations. This can be achieved in principle by
controlling local magnetic fields or local Overhauser fields. 
For a structure designed to apply ESR excitation to a single quantum dot,
see figure~\ref{Fig: localESR}. Another
approach is the individual control of the electron $g$-factor instead
of the local magnetic field. In quantum wells, there has
been  recent pioneering work in this direction 
\cite{salis:2001a,kato:2003a,kato:2004a}.
Salis {\it et al.} \cite{salis:2001a} have demonstrated electrically controlled
modulation of the $g$-factor in an AlGaAs quantum well containing
a gradient in the Al concentration. Here, the electron wave function
was shifted between regions with different Al concentration via applied
gate voltages, which resulted in the observation of a different electron
$g$-factor. Kato {\it et al.} \cite{kato:2003a} have even demonstrated 
voltage-controlled modulation of the $g$-tensor. This allows
the induction of ESR without time-dependent magnetic fields. Further
experiments by Kato {\it et al.} \cite{kato:2004a} exploited the spin-orbit
interaction to achieve coherent spin manipulation in strained semiconductor
films without the application of magnetic fields. %

Alternative proposals to produce single-spin rotations are related
to all-optical Raman transitions \cite{imamoglu:1999a} and stimulated 
Raman adiabatic passage (STIRAP) 
\cite{imamoglu:1999a,troiani:2003a,chen:2003a,calarco:2003a},
a method based on two-photon Raman transitions which has already been
applied to atoms and molecules to transfer a precisely controlled
 population between two quantum states \cite{bergmann:1998a}.
While Troiani {\it et al.} \cite{troiani:2003a} have also considered the realization
of conditional and unconditional quantum gates using an additional
adjacent quantum dot, Chen {\it et al.} \cite{chen:2003a} have proposed a STIRAP
process with no auxiliary state, but in the presence of a transverse
magnetic field. In this setup, control of the relative phase and the relative
intensity of two applied laser pulses enables an arbitrary spin rotation
for a given polarization of the light and direction of the transverse
magnetic field \cite{chen:2003a}. As an alternative method of performing a 
spin rotation on an excess electron confined to a quantum dot, Calarco 
{\it et al.} \cite{calarco:2003a}
have proposed to excite lh states via a sequence of a 
linearly and then a circularly polarized laser $\pi$-pulse. 
 Given this abundance of proposals for single-qubit
gates, there is great hope for working experimental realizations in
the near future.

\subsection{Two-Qubit Gates\label{sub:Two-Qubit-Gates}}

Swapping of the spin states of two electrons located in closely spaced
quantum dots seems by now to be a realistic first experimental step
towards a two-qubit gate for spins. As explained in section \ref{sec:Loss-DiVincenzo},
this can be achieved by controlling the overlap of the two wave functions
of the electrons and thus the singlet-triplet splitting $J$.
The interdot tunnel splitting and $J$ can be determined from a transport
experiment in the sequential tunnelling regime 
\cite{sukhorukov:1998a,sukhorukov:2001a,golovach:2004b}.
Recently, $J$ has been measured for two electrons in a single gated
quantum dot by detecting inelastic cotunnelling above and below a magnetic
field driven singlet-triplet transition \cite{zumbuhl:2004a}. 
In the cotunnelling regime, only second-order tunnelling processes 
contribute to charge transport. Because
the dot was elliptical, a two-electron wave function similar
to that in a double dot was expected. Two different samples yielded $J\approx0.2\:\mathrm{meV}$
and $J\approx0.57\:\mathrm{meV}$ at $B=0$. The critical
magnetic field for the singlet-triplet transition (where $J=0$) has been measured
to be $B^{*}\approx1.3\:\mathrm{T}$. For the interaction parameter 
\cite{golovach:2004b},
$\phi\approx0.5\pm0.1$ has been obtained, indicating that the ground state
given by 
$|+\uparrow,+\downarrow\rangle -\phi|-\uparrow,-\downarrow\rangle /\sqrt{1+\phi^2}$ (where $\pm$ stands for the symmetric/antisymmetric orbital wave function)
consists of a singlet with a significant admixture of single-electron orbitals
due to the electron-electron interaction. The entanglement of the two 
electron spins in the state above can be quantified  by the concurrence 
$C=2\phi /(1+\phi^2)$ \cite{golovach:2004b,schliemann:2001a}.  
The experimental result $C\approx 0.8$ shows that electron-electron 
interaction reduces the degree of spin
entanglement from its maximum ($C=1$), which is obtained for a singlet 
(having $\phi=1$). This demonstration strongly encourages that  similar 
results might be soon obtained in double dots (which are needed for 
spatially separating the two qubits).

In addition to the  two-qubit gate that is controlled via the tunnel 
coupling of the two dots (see section \ref{sec:Loss-DiVincenzo}), there is also a
proposal for an optical two-qubit phase gate \cite{calarco:2003a}.
In this proposal, a two-qubit phase gate is established by applying an 
adiabatically
chirped laser pulse (this is a pulse with a time-dependent frequency) to 
two neighbouring quantum dots, each with one excess
electron. The desirable phase of the two-qubit gate is accumulated during 
the (electrostatic) interaction
time of the two charged excitons that are excited in the two dots.
The adiabatic change of the laser detuning protects
 the system from interaction with phonons, even in the presence
of hole-state mixing. The combination of such a two-qubit
gate with an optical single-qubit gate (as outlined in section \ref{sub:Single-Qubit-Rotations})
would finally enable all-optical quantum computation using spins in
quantum dots.

\section{Conclusions}

In this tutorial we have discussed theoretical concepts and the present
status of experimental achievements towards the implementation of
quantum information processing using electron spins in quantum dots.
The demonstration of working single and two-qubit gates and finally
the production of quantum dot arrays that enable the application of
an entire quantum algorithm including error correction are the major
problems to tackle towards the goal of a solid-state implementation
of quantum information processing using electron spins in quantum
dots.

\section*{Acknowledgments}
We thank 
D N Bulaev, H-A Engel, K Ensslin, S I Erlingsson, V N Golovach,
A H\"ogele, L P Kouwenhoven, J Lehmann, C M Marcus, P M Petroff and W G van
der Wiel for useful discussions and for supplying valuable material to this
work. We acknowledge support from the NCCR Nanoscience, the Swiss NSF, NSERC of
Canada, EU RTN Spintronics, EU RTN QUEMOLNA, DARPA, ARO, and ONR.


\end{document}